\documentclass[traditabstract]{aa}

\usepackage{graphicx,float,lscape}
\usepackage{txfonts}
\begin{document}

   \title{VIRTIS-H observations of comet 67P's dust coma: spectral properties and color temperature variability with phase and elevation.}


   \author{D. Bockel\'ee-Morvan
          \inst{1}
          \and
          C. Leyrat \inst{1}
          \and
          S. Erard \inst{1}
          \and 
          F. Andrieu \inst{1} 
                   \and 
          F. Capaccioni \inst{2}
          \and G. Filacchione \inst{2}                     
          \and P.H. Hasselmann \inst{1}
          \and  J. Crovisier \inst{1}
          \and P. Drossart \inst{1}
                    \and G. Arnold \inst{3}
                    \and M. Ciarniello \inst{2} 
                    \and D. Kappel \inst{3}  
                   \and A. Longobardo \inst{2} 
                   \and M.-T. Capria \inst{2}
                   \and M.C. De Sanctis \inst{2}                                 
          \and G. Rinaldi \inst{2} 
          \and F. Taylor  \inst{4}                          
          }

   \institute{LESIA, Observatoire de Paris, PSL Research University, CNRS, Sorbonne Universit\'e, 
Univ. Paris-Diderot, Sorbonne Paris Cit\'e, \\ 5 place Jules Janssen, 92195 Meudon, France\\
              \email{dominique.bockelee@obspm.fr}
         \and
INAF-IAPS, Istituto di Astrofisica e Planetologia Spaziali, via del fosso del Cavaliere, 100, 00133, Rome, Italy   
		\and
Institute for Planetary Research, Deutsches Zentrum f\"{u}r Luft- und Raumfahrt (DLR), Rutherfordstrasse 2, 12489, Berlin, Germany 
\and
Departement of Physics, Oxford University, Parks Rd, OX13PJ, Regno Unito, Oxford, UK       
             }

   \date{Received ; accepted }


  \abstract{
We analyze 2--5 $\mu$m spectroscopic observations of the dust coma of comet 67P/Churyumov-Gerasimenko obtained with the VIRTIS-H instrument onboard Rosetta from
3 June to 29 October 2015 at heliocentric distances $r_{\rm h}$ = 1.24--1.55 AU. The 2--2.5 $\mu$m color, bolometric albedo, and color temperature are measured using spectral fitting. Data obtained at $\alpha$ = 90$^{\circ}$ solar phase angle show an increase of the bolometric albedo (0.05 to 0.14) with increasing altitude (0.5 to 8 km), accompanied by a possible marginal decrease of the color and color temperature.  Possible explanations include the presence in the inner coma of dark particles on ballistic trajectories, and radial changes in particle composition. 
In the phase angle range 50--120$^{\circ}$, phase reddening is significant (0.031 \%/100 nm/$^\circ$), for a mean color of 2 \%/100 nm at $\alpha$ = 90$^{\circ}$, that can be related to the roughness of the dust particles. Moreover, a decrease of the color temperature with decreasing phase angle is also observed at a rate of $\sim$ 0.3 K/$^{\circ}$, consistent with the presence of large porous particles, with low thermal inertia, and showing a significant day-to-night temperature contrast.
Comparing data acquired at fixed phase angle ($\alpha$ = 90$^{\circ}$), a 20\% increase of the bolometric albedo is observed near perihelion. 
Heliocentric variations of the dust color are not significant in the analyzed time period. Measured color temperatures are varying from 260 to 320 K, and follow a $r_{\rm h}^{0.6}$ variation in the  $r_{\rm h}$ = 1.24--1.5 AU range, close to the expected $r_{\rm h}^{0.5}$ value.

}


\titlerunning{67P dust coma}
\authorrunning{Bockel\'ee-Morvan et al.}
   \maketitle
%


\section{Introduction}
The Rosetta mission of the European Space Agency accompanied comet 67P/Churyumov-Gerasimenko (hereafter 67P) between 2014
and 2016 as it approached perihelion (13 August 2015) and receded from the Sun.
Several in situ intruments on the Rosetta orbiter were dedicated to the study of the physical and chemical properties of the dust particles released in the coma. 
The Micro-Imaging Dust
Analysis System \citep[MIDAS,][]{Riedler2007} acquired 
the  3D  topography of 1 to 50 $\mu$m  sized  dust particles  with  resolutions 
down to a few nanometers, and showed that dust particles are agglomerates at all scales with the smallest subunit sizes of less than 100 nm \citep{Bentley2016}. A highly porous fractal-like aggregate with a fractal dimension $D_f$ = 1.7 was collected \citep{Mannel2016}. The Cometary Secondary Ion Mass Analyzer \citep[COSIMA,][]{Kissel2007} collected dust particles to image them at a resolution of 14 $\mu$m and performed secondary ion mass spectroscopy.
Both porous aggregates and more compact particles were observed \citep{Langevin2016, Merouane2016}. The chemical analysis indicates that these particles are made of 50\% organic matter in mass, mixed with mineral phases that are mostly anhydrous \citep{Bardyn2017}. Carbon is mainly present as macromolecular material and shows similarities with the Insoluble Organic Matter (IOM) found in carbonaceous chondrites \citep{Fray2016}. The Grain Impact Analyzer and Dust Accumulator \citep[GIADA,][]{Colangeli2007} measured the scattered light, speed, and momentum of individual particles in the size range of typically 150--800 $\mu$m. The majority of the detected dust is described to be porous agglomerates with a mean density of 785$^{+520}_{-115}$ kg m$^{-3}$ \citep{Fulle2017}. GIADA also detected very low density, fluffy agglomerates, with properties similar to the MIDAS fractal particles \citep{Fulle2016}.  

The remote sensing instruments onboard Rosetta provide complementary information on the dust properties by measuring scattered light or thermal emission from  particles. From multi-color imaging using the Optical, Spectroscopic, and Infrared Remote Imaging System \citep[OSIRIS,][]{Keller2007}, spectral slopes were measured both for individual particles \citep{Frattin2017} and for the unresolved dust coma \citep{Bertini2017}. The observed reddening \citep[e.g. typically 11--14 \%/100 nm at $\lambda$ = 0.4--0.7 $\mu$m for the diffuse coma,][]{Bertini2017} is characteristic of  particles made of absorbing material, e.g. organics \citep[][and references therein]{Kolokolova2004}. Individual grains (sizes in the range of centimeters to decimeters) display differing spectra, which may be related to variations of the organic/silicate ratio and the presence of ice \citep{Frattin2017}. The spectral slopes measured on individual grains display variations with heliocentric and nucleocentric distances that could be related to physical processes in the coma  affecting  the released material \citep{Frattin2017}. However, spectrophotometric data of the diffuse coma obtained with OSIRIS do not show any trend with heliocentric distance and nucleocentric distance \citep{Bertini2017}.

In this paper, we analyze 2--5 $\mu$m spectra of continuum radiation from the dust  coma acquired with the high spectral resolution channel of the Visible InfraRed Thermal Imaging Spectrometer (VIRTIS-H) onboard Rosetta \citep{Coradini2007}. This paper is a follow-up of previous work published by \citet{Rinaldi2016,Rinaldi2017} (VIRTIS-M data) and \citet{dbm2017} (VIRTIS-H data). Whereas those studies  provided information on the scattering and thermal properties of 67P's quiescent dust coma at a few dates, namely March-April 2015 and September 2015, we analyze here a comprehensive set of VIRTIS-H data acquired from June to October 2015 (encompassing perihelion on 13 August 2015), with the heliocentric distance spanning $r_{\rm h}$ = 1.24--1.55 AU. We  derive the bolometric albedo and color temperature of the dust coma, as well as the spectral slope between 2--2.5 $\mu$m following \citet{Gehrz1992}. These parameters, which have been measured for several comets, depend on the size distribution, porosity, and composition of the dust particles \citep{Kolokolova2004}.  Measurements obtained one month after perihelion at $r_{\rm h}$ = 1.3 AU and 90$^{\circ}$ phase angle are consistent with values measured for most comets \citep{dbm2017}. In this paper, we seek for possible variations with heliocentric distance, altitude, and phase angle.

Section \ref{sec:Sect1} presents the data set. The spectral analysis is described in Sect.~\ref{sec:modelfitting}. Results are given in Sect.~\ref{sec:results}. A discussion on the observed trends with phase angle and altitude follows in Sect.~\ref{sec:discussion}. Appendix~\ref{sec:appendixB} presents expected thermal properties of dust particles, and the model used to interpret the variation of the color temperature with phase angle.

\section{The VIRTIS-H data set}
\label{sec:Sect1}

VIRTIS is composed of two
channels: VIRTIS-M, a spectro-imager with a visible (VIS)
(0.25--1 $\mu$m) and an infrared (IR)  (1--5 $\mu$m) channel operating at moderate
spectral resolution ($\lambda$/$\Delta \lambda$ = 70-380), and
VIRTIS-H, a cross-dispersing spectrometer providing spectra with higher spectral resolution capabilities
($\lambda$/$\Delta \lambda$ = 1300-3500) in eight
orders of diffraction covering the range 1.9--5.0 $\mu$m (Table~\ref{tab:order}) \citep{Drossart2000,Coradini2007}. The infrared channel of VIRTIS-M underwent a cryocooler failure at the beginning of May 2015. After this date, infrared data were only collected with VIRTIS-H, and we are focusing on these data.

As for most Rosetta instruments, the line of sight of VIRTIS-H is along the Z-axis of the spacecraft (S/C). The instantaneous field of view (FOV) of this
point instrument is 0.58 $\times$ 1.74 mrad$^2$ (the larger
dimension being along the Y axis). Details on the calibration process are given in \citet{dbm2016}. The version of the calibration pipeline is CALIBROS--1.2--150126.

VIRTIS-H acquired data cubes of typically 3 h duration in various pointing modes. For coma studies, the main observing modes were: 1) limb sounding at a given distance from the comet surface along the comet-Sun line; 2) limb sounding at a few stared positions along the comet-Sun line; 3) limb sounding at a few altitudes  and azimuthal angles with respect to the comet-Sun direction; 4) raster maps  \citep[see examples in][]{dbm2016}. The data used in this paper were obtained with pointing modes 1--3. Dust continuum maps obtained from rasters will be the topic of a future paper.   

\begin{figure}
    \includegraphics[width=\columnwidth]{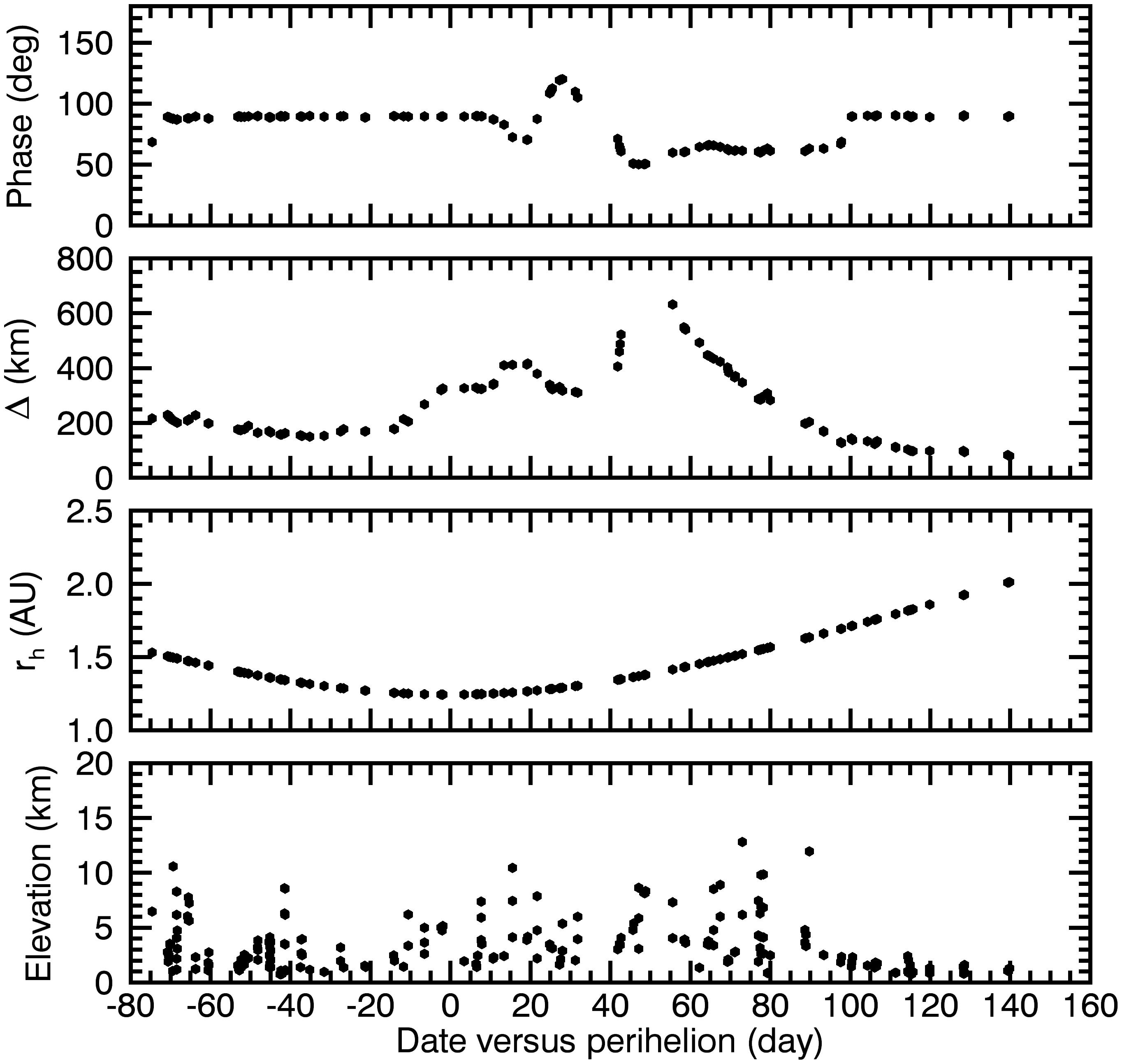}
       \caption{Geometrical informations for the constituted data set of averaged limb spectra. From top to bottom: S/C-nucleus-Sun angle (referred as to the phase angle), S/C distance to comet center ($\Delta$), heliocentric distance ($r_{\rm h}$), and mean elevation ($H$) of the line-of-sight.   }
    \label{fig:geo}
\end{figure}

We considered data cubes acquired from MTP016/STP058  to MTP024/STP089 covering dates from 30 May 2015 ($r_{\rm h}$ = 1.53 AU) to 30 December 2015 ($r_{\rm h}$ = 2.01 AU), that is, from 74 days before perihelion to 139 days after perihelion.
In total 141 data cubes were used, though those acquired after 29 October 2015  turned out to be not appropriate for model fitting of the dust continuum due to low signals (see below). Spectra were obtained by co-adding acquisitions in the coma for which the exposure time was typically 3 s. Since we were interested in studying whether spectral characteristics vary with nucleus distance, we co-added acquisitions by ranges of tangent altitude (hereafter referred as to the elevation) with respect to the nucleus surface. This was done when the signal-to-noise ratio was high enough, and when the elevation significantly varied during the acquisition of the data cube (i.e., for pointing modes 2--3). In total 222 spectra were studied.  Figure~\ref{fig:geo}
provides information for each of these spectra regarding the heliocentric distance, the S/C distance to nucleus center ($\Delta$), and the S/C-nucleus-Sun angle (referred to as the phase angle). The mean elevation for these spectra is between 0.8 to 21 km, with 64\% of the spectra in the 0.8--4 km range, and 30\% of the spectra in the 4--10 km range (Fig.~\ref{fig:geo}). For 83\% of the spectra, the co-added acquisitions where taken at elevations which differ by less than 0.5 km. For the rest of the spectra, elevations of individual acquisitions differ by less than 1.5 km. For stared limb pointing, variation of elevation with time is observed due to the mutual effects of the complex shape of the 67P rotating nucleus and S/C motion. To define the elevation at which the spectra refer to, we took the weighted mean of the elevation value of each acquisition, with the weight equal to 1/$\rho$, where $\rho$ is the distance to nucleus center and is taken equal, for simplicity, to the elevation plus the mean radius of 67P nucleus of 2 km. Indeed column densities are expected to vary with a law close to 1/$\rho$, so we expect a larger contribution to the signal from acquisitions with a line-of-sight  closer to the nucleus.

Since the VIRTIS-H faint coma signals are affected by stray light coming from the nearby nucleus, a specific strategy has been implemented to manage these effects.
Stray light polluted the low wavelength range of each order, and more significantly order numbered 0, covering the 4--5 $\mu$m range (Table~\ref{tab:order}). Data cubes obtained at small elevations are the most affected by stray light. An algorithm  developed for stray light removal was applied (Andrieu et al., in preparation). However, in some cases, the algorithm was not able to remove all the stray light, especially in order 0. Therefore, the different orders (which overlap in wavelength coverage) were merged by selecting the sections of the orders which are not significantly affected by stray light. The selected wavelength ranges for each order (Table~\ref{tab:order}) allow to reassemble the entire spectrum in the 2--5 $\mu$m range. However, the 4.2--4.5 $\mu$m section of order 0 is affected by stray light but also by the presence of CO$_2$ fluorescence emissions, so it will not be considered for the analysis of the dust continuum radiation.  The degree of stray light pollution was estimated by computing the excess of radiance in order 0 with respect to order 1 at wavelengths where these orders overlap ($\sim$ 4.2 $\mu$m). We excluded spectra where this excess is larger than 40\%.

At the junction of the selected ranges of the different orders, spectra with low signal-to-noise ratio show intensity discrepancies to varying degrees, due to the instrumental response which is low on the edges of each order.  These defects were found to lead to inaccurate results when performing model fitting of the dust continuum radiation. We defined a criterion based on the ratio of the flux measured at 3 $\mu$m in order 4 with respect to the value measured at the same wavelength in order 3. In the initial sample of 222 spectra, this ratio, referred as to $TEST_{3.0}$, varies between 0.95 to 2.7, with a value close to 1 indicating a high quality spectrum. We only considered model fitting results for spectra complying $TEST_{3.0}$ $<$ 1.35 (173 spectra among the 222). In the following sections, we will also discuss results obtained for the best quality spectra fulfilling $TEST_{3.0}$ $<$ 1.1 (49 spectra). For these high-quality spectra, the signal-to-noise ratio (SNR) at 4.65 $\mu$m is in the range 30--80, with a few exceptions (the relevant root mean square is computed on the spectrum from the statistics of the residuals between the observed spectrum and model fit (Sect. \ref{sec:modelfitting}) in the range 4.5--4.8  $\mu$m). The SNRs at 3.3 $\mu$m (order 3) and 2.3 $\mu$m (order 7) are a factor 3--4 lower. Spectra with SNR at 4.65 $\mu$m less than 12 were not considered. After excluding spectra with high stray light pollution, at the end, 99 (respectively 49) spectra complying $TEST_{3.0}$ $<$ 1.35 (respectively $TEST_{3.0}$ $<$ 1.1) where found appropriate for model fitting.  The covered time period is --71 to +78 d with respect to perihelion (3 June to 29 Oct. 2015, $r_{\rm h}$ = 1.24 to 1.55 AU). The best quality spectra cover dates from --43 to +78 d. Table~\ref{tab:logtable} provides information on these 99 spectra, such as VIRTIS-H observation identification number, start time of the data cube, date with respect to perihelion, spacecraft distance to nucleus center, heliocentric distance and phase angle.

\begin{figure}
    \includegraphics[width=\columnwidth]{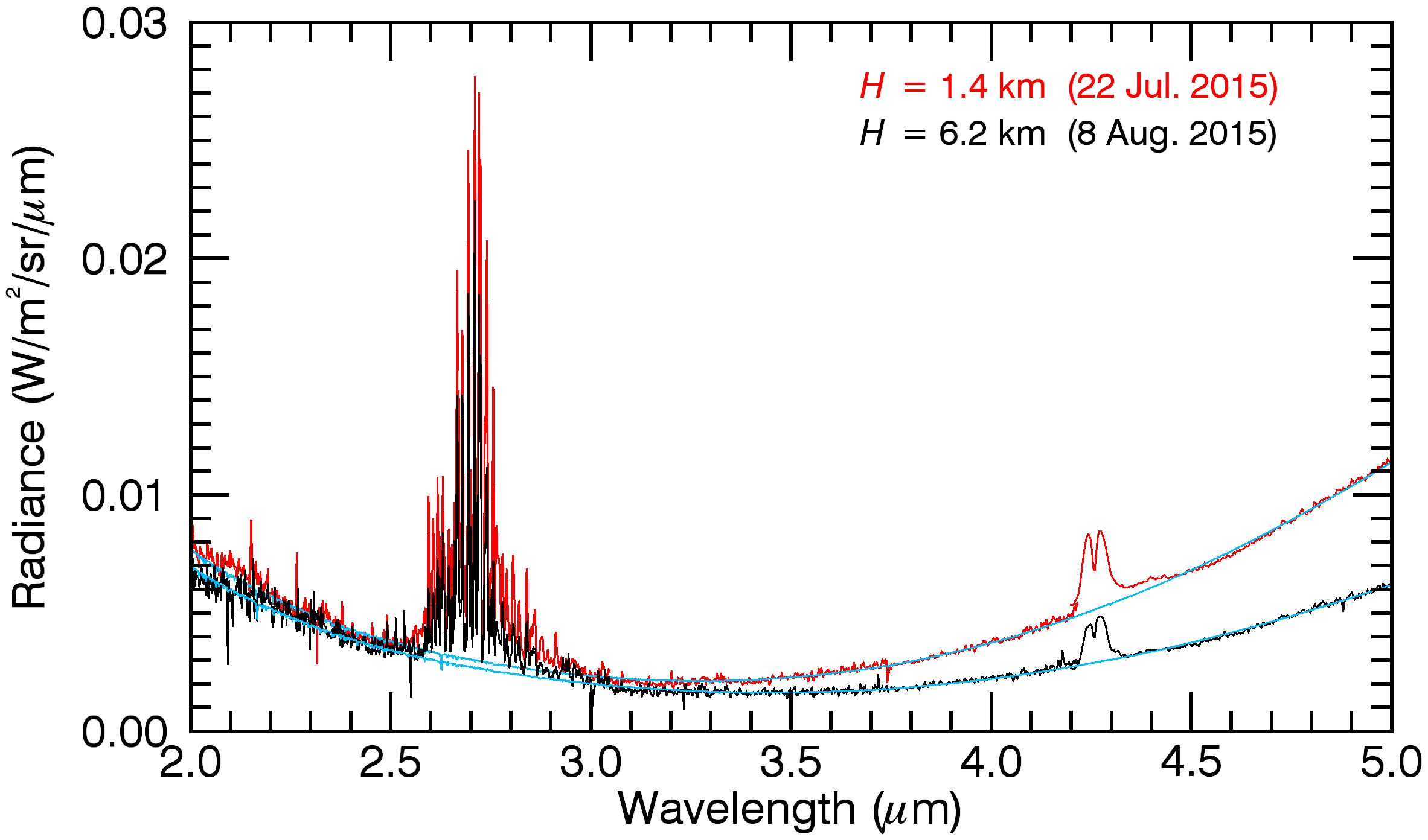}
       \caption{VIRTIS-H spectra of comet 67P obtained at different elevations above the surface. Cube T1$\_$00396220410 acquired on 22 Jul. 2015 ($r_{\rm h}$ = 1.27 AU) with a mean elevation $H$ = 1.4 km (red line). Cube T1$\_$00397139303  acquired on 8 Aug. 2015 ($r_{\rm h}$ = 1.25 AU) using selected acquisitions with  $H$ = 6.2 km (black line). The phase angle is $\sim$89$^\circ$ for both spectra. The model fits to the continuum are shown in cyan, with derived parameters ($T_{\rm col}$(K), $A$, $S'_{\rm col}$(\% per 100 nm)) = (295, 0.07, 2.3) and (289, 0.10, 2.2) for the 22 Jul. and 8 Aug. spectra, respectively. The spectra fulfill the quality criterion $TEST_{3.0}$ $<$ 1.1.  }
           \label{fig:splowhighalbedo}
\end{figure}

Figure~\ref{fig:splowhighalbedo} shows two examples of high quality spectra affected by negligible stray light, obtained for the coma on 22 Jul. 2015 and 8 Aug. 2015 with SNR of respectively 76 and 50 at 4.65 $\mu$m. The dust continuum consists of  scattered sunlight at $\lambda$ < 3.5 $\mu$m and thermal radiation at longer wavelengths. Fluorescence emission bands of H$_2$O, CO$_2$ and $^{13}$CO$_2$ are observed in the 2.5--3.0 $\mu$m and 4.2--4.4 $\mu$m ranges \citep[see detailed description in][]{dbm2015,dbm2016}.

\begin{table}
\caption{VIRTIS-H diffraction orders and selected wavelengths.\label{tab:order}}
\begin{tabular}{l l l }        
\hline\hline \noalign{\smallskip}
order & $\lambda$ coverage & selected $\lambda$ range \\
\cline{2-3}\noalign{\smallskip}
& \multicolumn{2}{c}{($\mu$m)} \\
\hline\noalign{\smallskip}
0 & 4.049--5.032 & 4.199--5.000$^a$\\
1 & 3.477--4.325 & 3.751--4.216\\
2 & 3.044--3.774 & 3.340--3.751\\
3 & 2.703--3.368 & 3.047--3.344\\
4 & 2.432--3.077 & 2.736--3.047\\
5 & 2.211--2.755 & 2.506--2.736\\
6 & 2.024--2.526 & 2.316--2.507$^b$\\
7 & 1.871--2.331 & 2.005--2.311$^c$\\
\hline \noalign{\smallskip}
\end{tabular}
{\footnotesize

$^a$ 4.2--4.5 $\mu$m range excluded for spectral fitting.

$^b$ 2.3--2.38 $\mu$m range excluded for spectral fitting.

$^c$ 2.08--2.17 $\mu$m range excluded for spectral fitting.
}

\end{table}

\section{Model fitting}  
\label{sec:modelfitting}

\begin{figure}
    \includegraphics[width=\columnwidth]{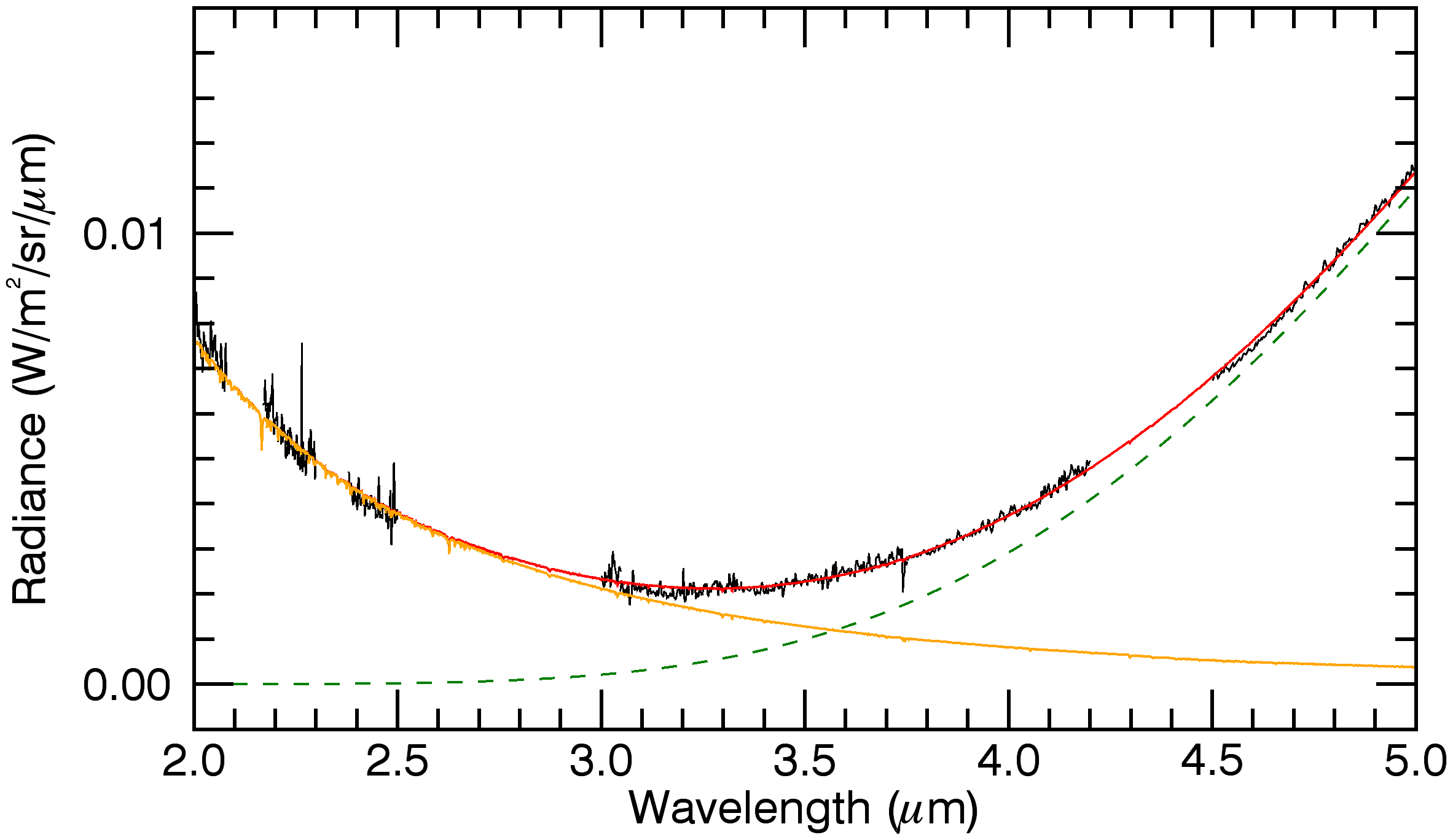}
       \caption{Example of model fit. The spectrum in black is a VIRTIS-H spectrum of comet 67P obtained on 22 Jul. 2015 (T1$\_$00396220410), with the regions excluded for fitting or presenting water and CO$_2$ emission lines not shown (the full spectrum is given in Fig.~\ref{fig:splowhighalbedo}). The model fit to the continuum, which corresponds to the sum of scattered light (plain orange line) and thermal radiation (dashed green line) is shown in red.  Retrieved parameters are $T_{\rm col}$ = 295 $\pm$ 1 K (corresponding to $S_{\rm heat}$ = 1.194 $\pm$ 0.003), $A$ = 0.068 $\pm$ 0.001,  $S'_{\rm col}$ = 2.3 $\pm$ 0.1 \% per 100 nm.     }
           \label{fig:sp-fit}
\end{figure}  

\begin{figure*}[t]
\centering
\begin{minipage}{8cm}
    \includegraphics[width=\columnwidth]{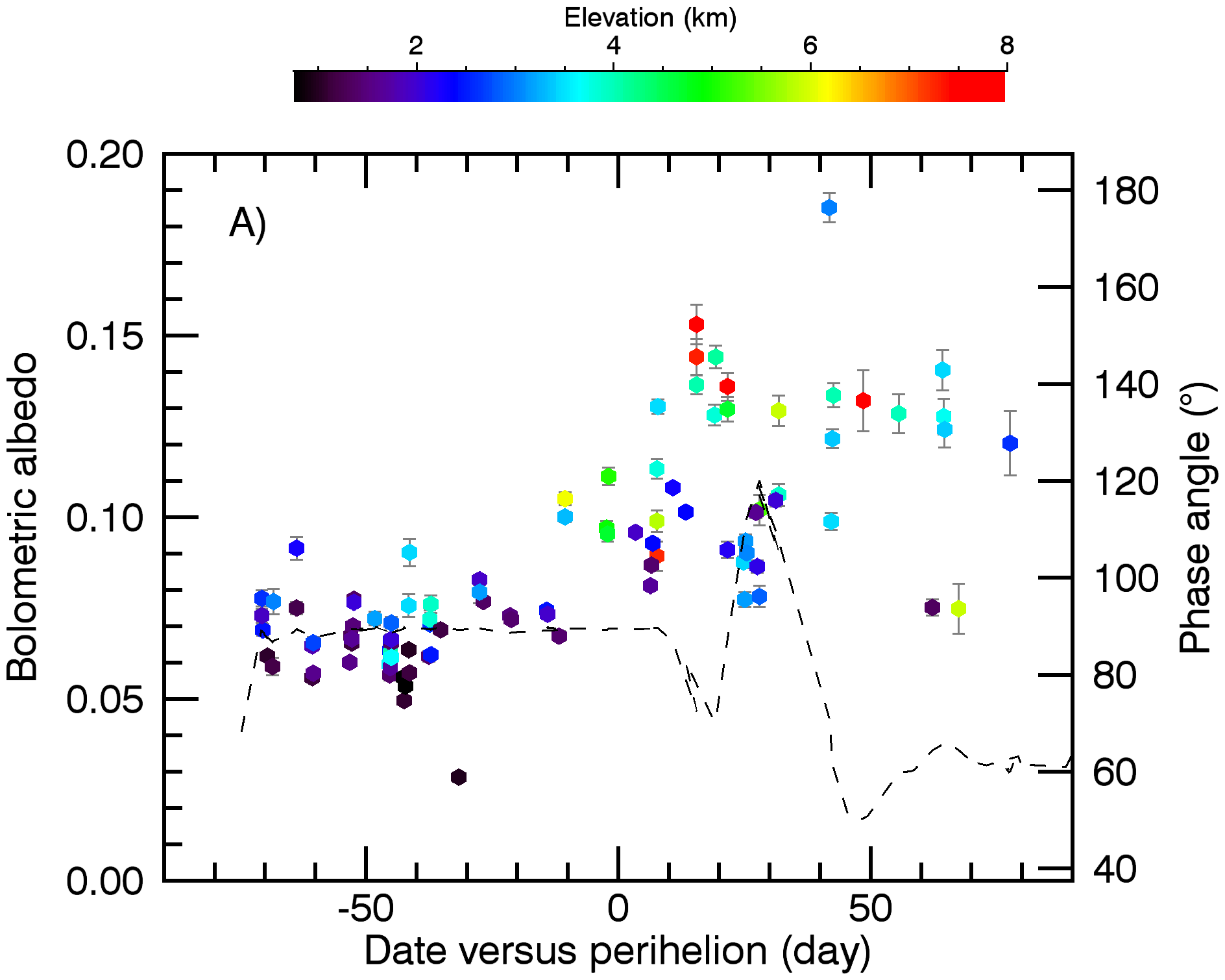}
\end{minipage}\hfill
\begin{minipage}{6.9cm}
\hspace{-1.05cm}
\vspace{-1.08cm}
    \includegraphics[width=\columnwidth]{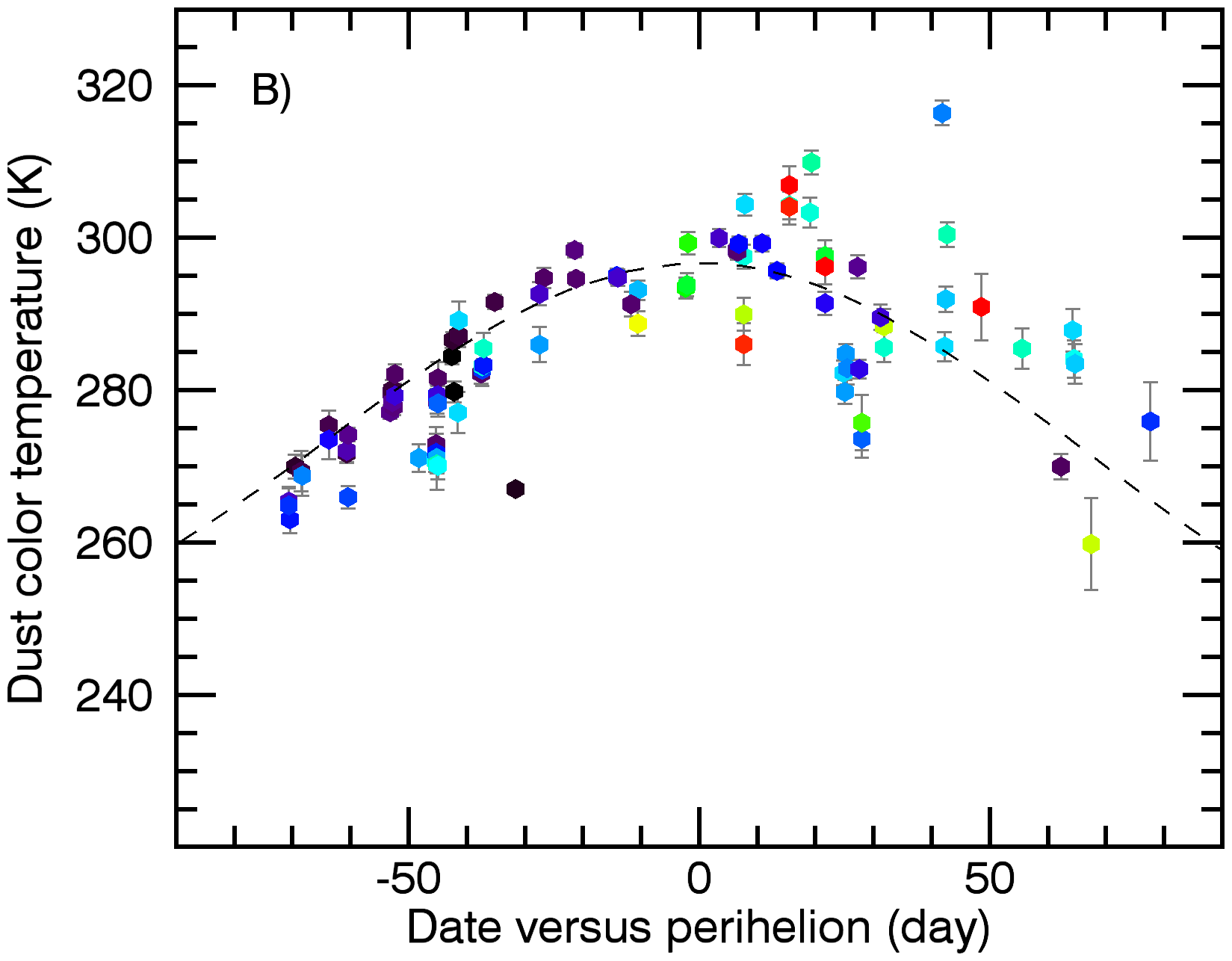}
\end{minipage}
\begin{minipage}{8.0cm}    
    \includegraphics[width=\columnwidth]{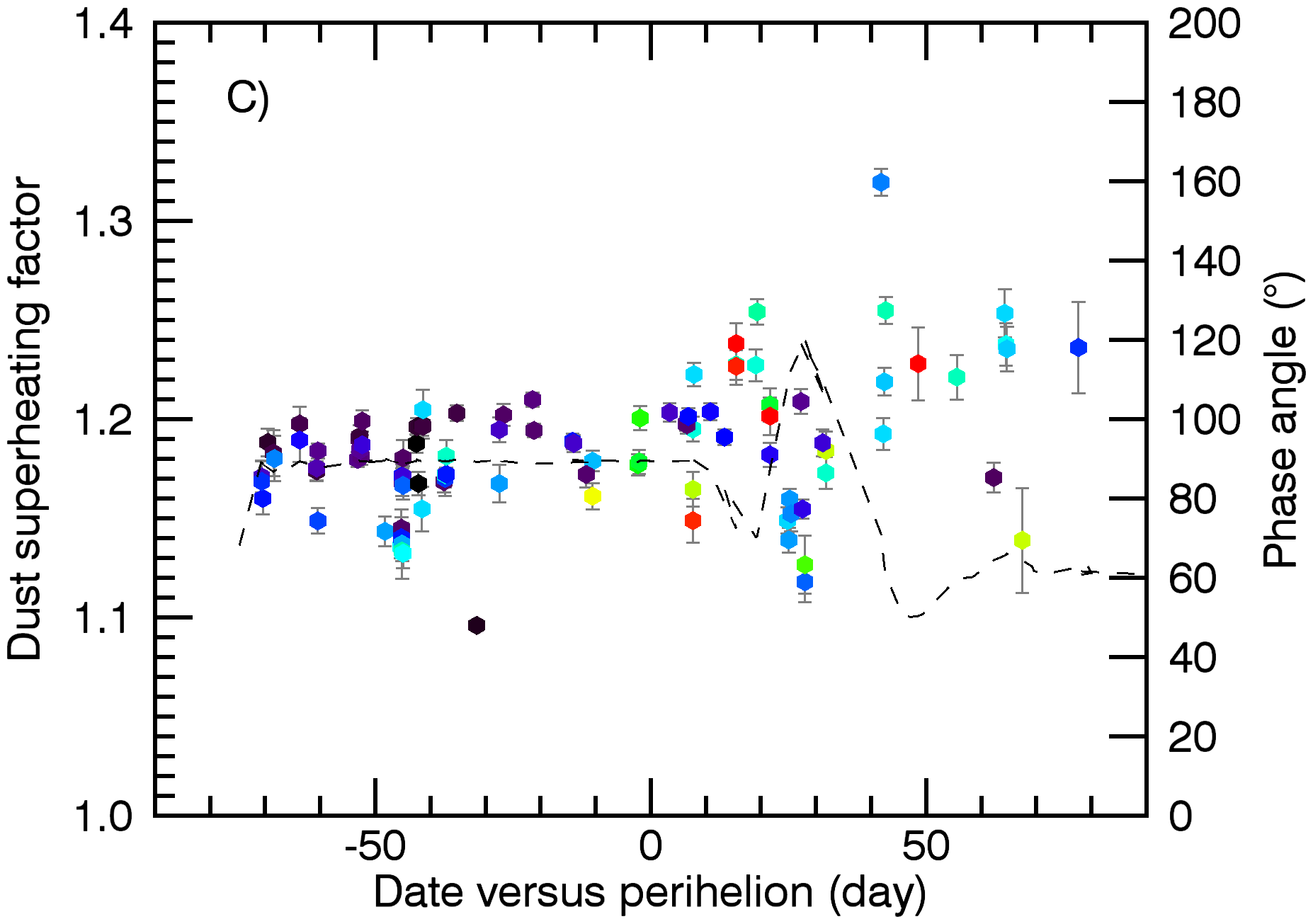}
\end{minipage}\hfill
 \begin{minipage}{8.0cm}    
    \includegraphics[width=\columnwidth]{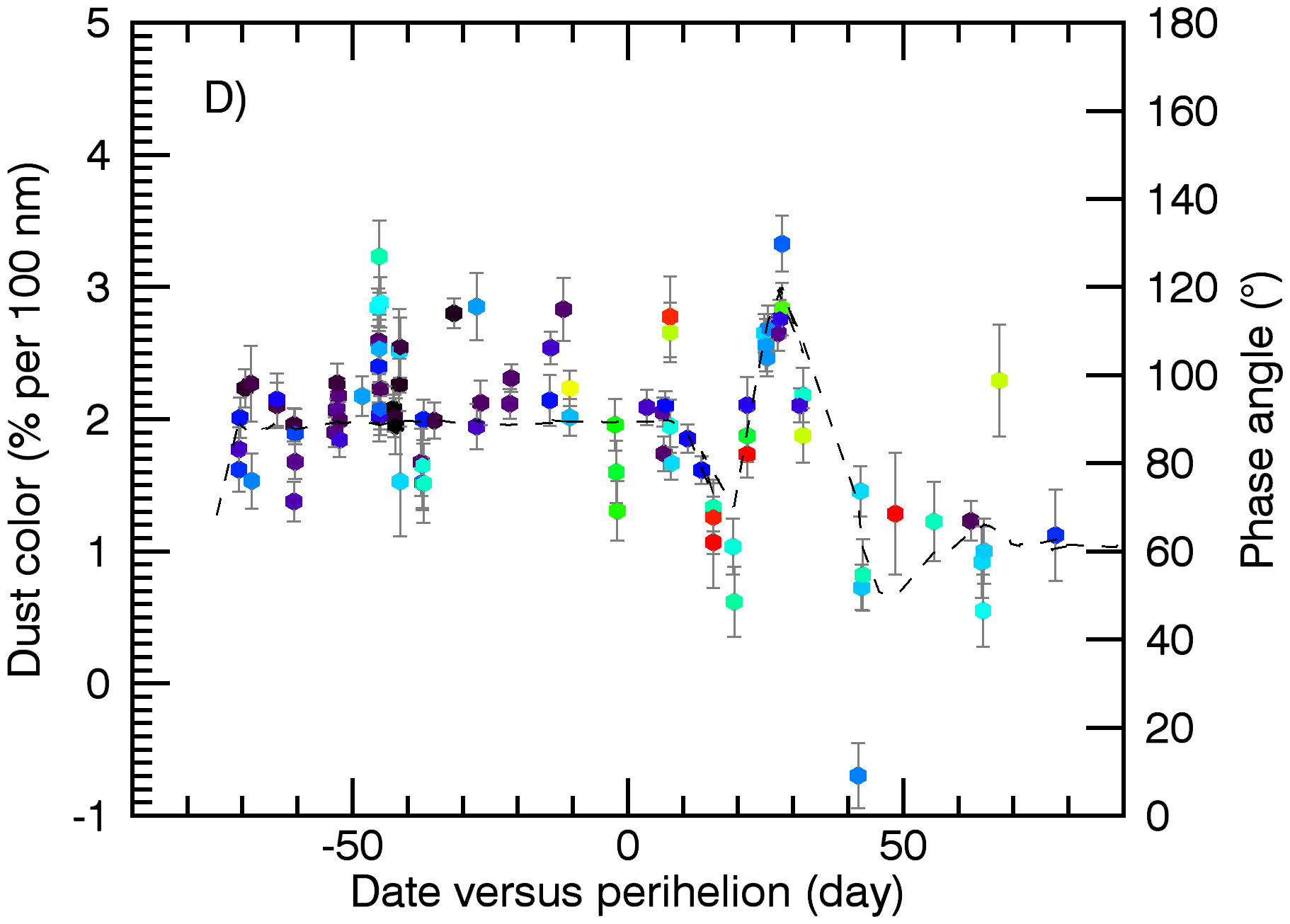}
\end{minipage}   
       \caption{Bolometric albedo (A), color temperature (B), superheating factor (C) and dust color (D) as a function of time with respect to perihelion. The color of the symbols is according to elevation (as indicated in the colorbar). The  phase angle is plotted as a dashed line in plots A, C, and D. The dashed line in plot B) is the color temperature obtained for a value of the superheating factor of 1.19 and a heliocentric variation following $r_{\rm h}^{-0.5}$ (Eq.~\ref{eq:1}). The fitted spectra fulfill the condition {\it TEST}$_{3.0}$ $<$ 1.35. } 
\label{fig:res_allphase}
\end{figure*}

In order to analyze the dust continuum radiation, we followed the approach presented by \citet{dbm2017} which consists in modeling the dust spectrum as the sum of scattered solar flux and thermal emission (described by a blackbody function). The free parameters of the model fitting are the color temperature $T_{\rm col}$, the spectral index of the reflectance, which allows us to derive the dust color $S'_{\rm col}$ in the 2.0--2.5 $\mu$m range, and the bolometric albedo $A$($\theta$), where $\theta$ is the scattering angle (hereafter we will use instead the phase angle $\alpha= 180^{\circ} - \theta$, and assimilate the phase angle to the S/C--Comet--Sun angle, which is a good approximation given the large S/C distance to the comet). From the color temperature, we can derive the so-called superheating factor $S_{\rm heat}$, defined as the ratio of the observed color temperature $T_{\rm col}$ to the equilibrium temperature $T_{\rm equ}$ of a fast rotating body:

\begin{equation}\label{eq:1}
S_{\rm heat} = \frac{T_{\rm col}}{T_{\rm equ}},
\end{equation}

\noindent
with

\begin{equation}\label{eq:Teq}
T_{\rm equ} = {278r_{\rm h}^{-0.5}} [K],
\end{equation}

\noindent
where $r_{\rm h}$ is in AU (this unit is used throughout the paper).

The definitions of $A$($\theta$) and $S_{\rm heat}$ follow the prescription of \citet{Gehrz1992}, which allowed us to compare 67P's dust infrared emission properties to other comets for which these parameters have been measured  \citep{dbm2017}. The bolometric albedo $A$($\theta$) is approximately equal to the ratio between the scattered energy by the coma to the total incident energy, and scales proportionally to the geometric albedo times the phase function. Further details can be found in \citet{dbm2017}.

The dust color (or reddening) is measured in \%/100 nm using the dust reflectance at 2.0 $\mu$m and 2.5 $\mu$m:

\begin{equation}\label{eq:5}
S'_{\rm col} = (2/500) \times \frac{R_{\rm scatt}^{\rm fit}(2.5\mu m)-R_{\rm scatt}^{\rm fit}(2.0\mu m)}{R_{\rm scatt}^{\rm fit}(2.5\mu m)+R_{\rm scatt}^{\rm fit}(2.0\mu m)},
\end{equation} 

\noindent
where $R_{\rm scatt}^{\rm fit}$($\lambda$) is the fitted scattered light (Fig.~\ref{fig:sp-fit}) at the wavelength $\lambda$ divided by the solar flux at $\lambda$ \citep{Kurucz1992}. 

In the fitting process, the spectral region 4.2--4.5 $\mu$m showing CO$_2$ emissions and stray light was masked. However, unlike in \citet{dbm2017}, the 3.3--3.6 $\mu$m region was kept, as only very faint emission features from organics are observed in this region \citep{dbm2016}. The model includes a synthetic H$_2$O fluorescence spectrum (described in \citet{dbm2015} with a rotational temperature of 100 K) with the total intensity used as free parameter, so the 2.5--3.0 $\mu$m region presenting water lines could be considered. Despiking (using median filtering) was applied, removing spikes such as those seen in the spectrum of Fig.~\ref{fig:sp-fit}, though this was not found critical. We checked that the fitting method, which uses the $\chi^2$ minimization algorithm of Levenberg-Marquardt, provides correct results by applying it to synthetic spectra to which synthetic noise resembling the noise present in 67P spectra was added.
Applying our algorithm to the data set presented in Sect.~\ref{sec:Sect1}, the  best fits have a reduced $\chi^2$ very close to 1 (0.94 on average).

Figure~\ref{fig:sp-fit} shows an example of a model fit to a high-SNR dust spectrum, with the two components, scattered light and thermal emission, shown separately, and the retrieved free parameters indicated in the caption. The uncertainties in the retrieved parameters are probably somewhat underestimated because they only consider statistical noise and not defects in the spectra related, e.g., to the calibration, or possible residual stray light (see Sect.~\ref{sec:Sect1}). 
For example, for the fit shown in Fig.~\ref{fig:sp-fit}, 1-$\sigma$ uncertainties  are 0.3\%, 1\%, and 4\% for  $T_{\rm col}$(K), $A$ and $S'_{\rm col}$, respectively \citep[1-$\sigma$ confidence levels were derived as explained in][]{dbm2017}. However, though the noise level is low between 4.5 and 5 $\mu$m (SNR = 76), the fit is not fully satisfactory in this spectral region (Fig.~\ref{fig:sp-fit}). There is also a small radiance offset at 3.752 $\mu$m, which corresponds to the junction of the selected wavelength ranges in orders 1 and 2 (Table~\ref{tab:order}).  

It is important to point out that the retrieved free parameters are somewhat correlated. This is because scattered light and thermal emission contribute both to the continuum in a significant fraction of the 2--5 $\mu$m spectrum (Fig.~\ref{fig:sp-fit}). A statistical analysis based on contours of equal $\chi^2$ shows that $T_{\rm col}$ and $S'_{\rm col}$ (and consequently $A$) are correlated among them. Dust color and color temperature are negatively correlated, whereas the bolometric albedo and color temperature are positively correlated. As a result, significant flaws somewhere in the spectrum can lead to spurious results which follow this trend (e.g., a lower $T_{\rm col}$ combined with higher  $S'_{\rm col}$, and lower $A$). Effectively, we observed that spectra fulfilling the quality test $TEST_{3.0}$ $>$ 1.1 have lower $T_{\rm col}$, combined with higher  $S'_{\rm col}$ and lower $A$, compared to values retrieved for higher quality spectra with $TEST_{3.0}$ $<$ 1.1. This will be further discussed in Sect.~\ref{sec:terminator}. 

\section{Results}  
\label{sec:results}

Figure~\ref{fig:res_allphase} shows the bolometric albedo, color, color temperature and superheating factor as a function of date with respect to perihelion, for the 99 spectra with $TEST_{3.0}$ $<$ 1.35 and stray light excess $<$ 1.4, as explained in Sect.~\ref{sec:Sect1}. The different points also characterize the  2--5 $\mu$m dust emission at various elevations of the line of sight (as indicated by the color code), and phase angles (cf overplotted phase information on Fig.~\ref{fig:res_allphase}A, C, D). We recall that the elevation $H$ corresponds to the altitude of the tangent point (Sect.~\ref{sec:Sect1}).
The results are also listed in Table~\ref{tab:logtable}. 

$T_{\rm col}$ ranges from 260 K to 320 K, and follows approximately the
 $r_{\rm h}^{-0.5}$ variation expected from the balance between absorbed solar radiation and radiated thermal energy (Fig.~\ref{fig:res_allphase}B). The superheating factor $S_{\rm heat}$ is typically 1.2 before perihelion (phase angle $\alpha$ of about 90$^{\circ}$). However, strong 
  variations of $S_{\rm heat}$ are observed after perihelion when the Rosetta S/C was flying out of terminator (with $\alpha$ on the order of 60$^{\circ}$ or reaching 120$^{\circ}$). These variations seem to be correlated with changes in the phase angle (Fig.~\ref{fig:res_allphase}C). A strong correlation with phase angle is also observed for the color $S'_{\rm col}$ (Fig.~\ref{fig:res_allphase}D). Whereas $S_{\rm heat}$ decreases with increasing phase angle, the reverse is observed for the color. As for the bolometric albedo, higher values are measured after perihelion (Fig.~ \ref{fig:res_allphase}A), which is consistent with the phase function of cometary dust, which has a "U" shape with a minimum at 
$\alpha$ = 90--100$^{\circ}$ \citep{Bertini2017}. However, a trend for higher albedos at higher elevations and/or near perihelion is also suggested (Fig.~ \ref{fig:res_allphase}A).

In the subsequent subsections, we will analyze elevation/time and phase variations of $T_{\rm col}$, $A(\theta)$ and $S'_{\rm col}$. We will also study the intensity ratio between scattered light and thermal emission.  The reference for  scattered light is the radiance measured at $\lambda$ = 2.44 $\mu$m, obtained from the median of the radiances between 2.38 and 2.5 $\mu$m (order 6, Table~\ref{tab:order}). For the thermal emission,  the reference is the radiance at  $\lambda$ = 4.6 $\mu$m (median of radiances between 4.5--4.7 $\mu$m). The intensity ratio $f_{\rm scatt}/f_{\rm therm}$ is obtained by ratioing radiances in units of W m$^{-2}$ sr$^{-1}$ Hz$^{-1}$. At constant phase angle, if the dust size distribution and composition do not vary with time and in the coma, this ratio is expected to only exhibit a heliocentric dependence proportional to $r_{\rm h}^{-2}$$/$$BB$($T_{\rm col}$), where $BB$ is the blackbody function at $T_{\rm col}$ which varies as $r_{\rm h}^{-0.5}$. We corrected the derived intensity ratios from this heliocentric dependence assuming $S_{\rm heat}$ = 1.2 and converted  
it to the value at 1 AU ($f_{\rm scatt}/f_{\rm therm}$(1AU)).
As discussed at the end of Sect.~\ref{sec:modelfitting}, spectral fitting to spectra presenting some offsets at the junction of the orders can provide inaccurate results.  On the other hand, the intensity ratio $f_{\rm scatt}/f_{\rm therm}$(1AU) is directly measured on the spectra, and provides reliable  trends.

\subsection{Results at 90$^\circ$ phase angle}  
\label{sec:terminator}

\begin{figure}
\centering
\begin{minipage}{8cm}
    \includegraphics[width=\columnwidth]{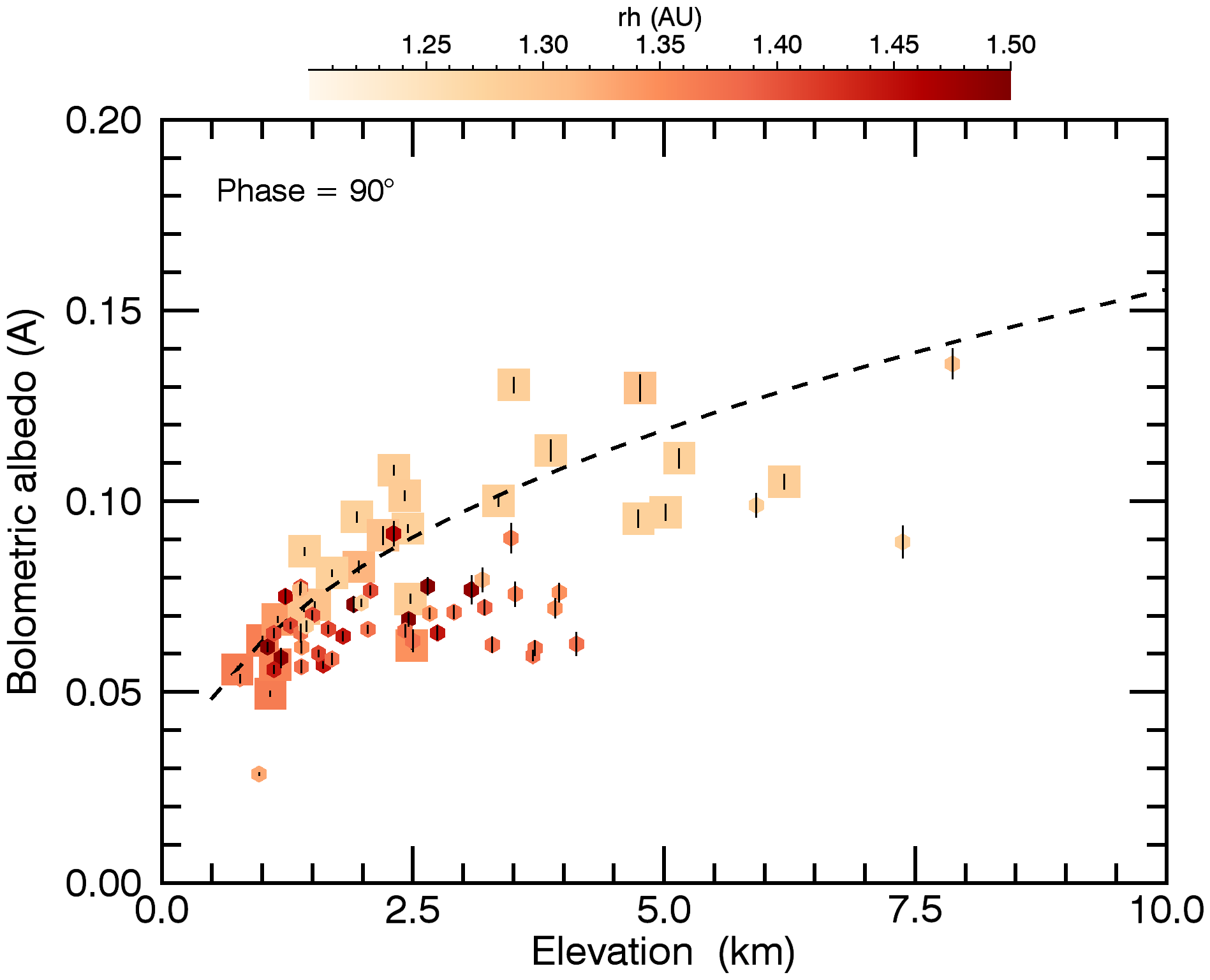}
\end{minipage}
\begin{minipage}{8cm}
    \includegraphics[width=\columnwidth]{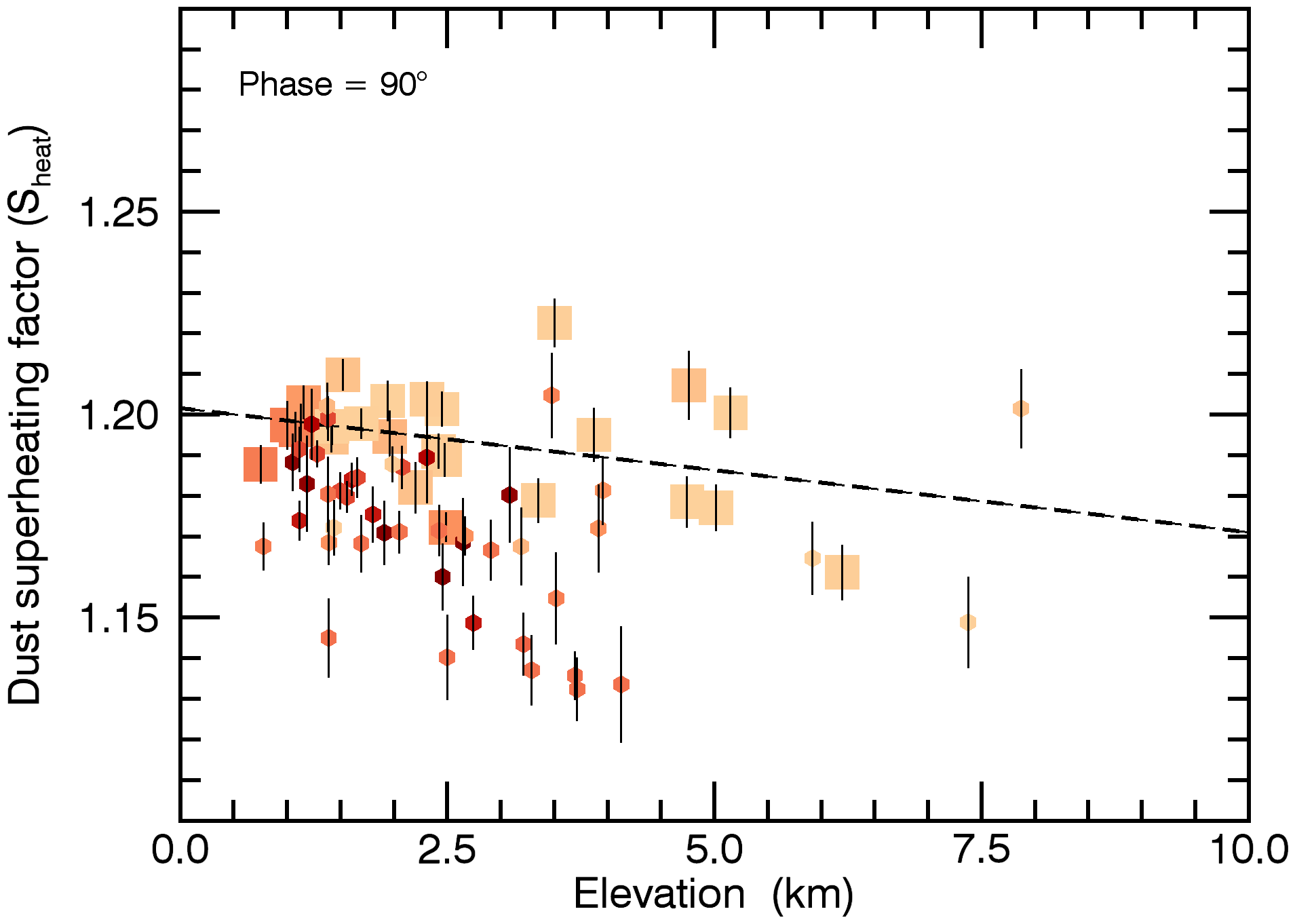}
\end{minipage}
\begin{minipage}{7.2cm}    
    \includegraphics[width=7.6cm]{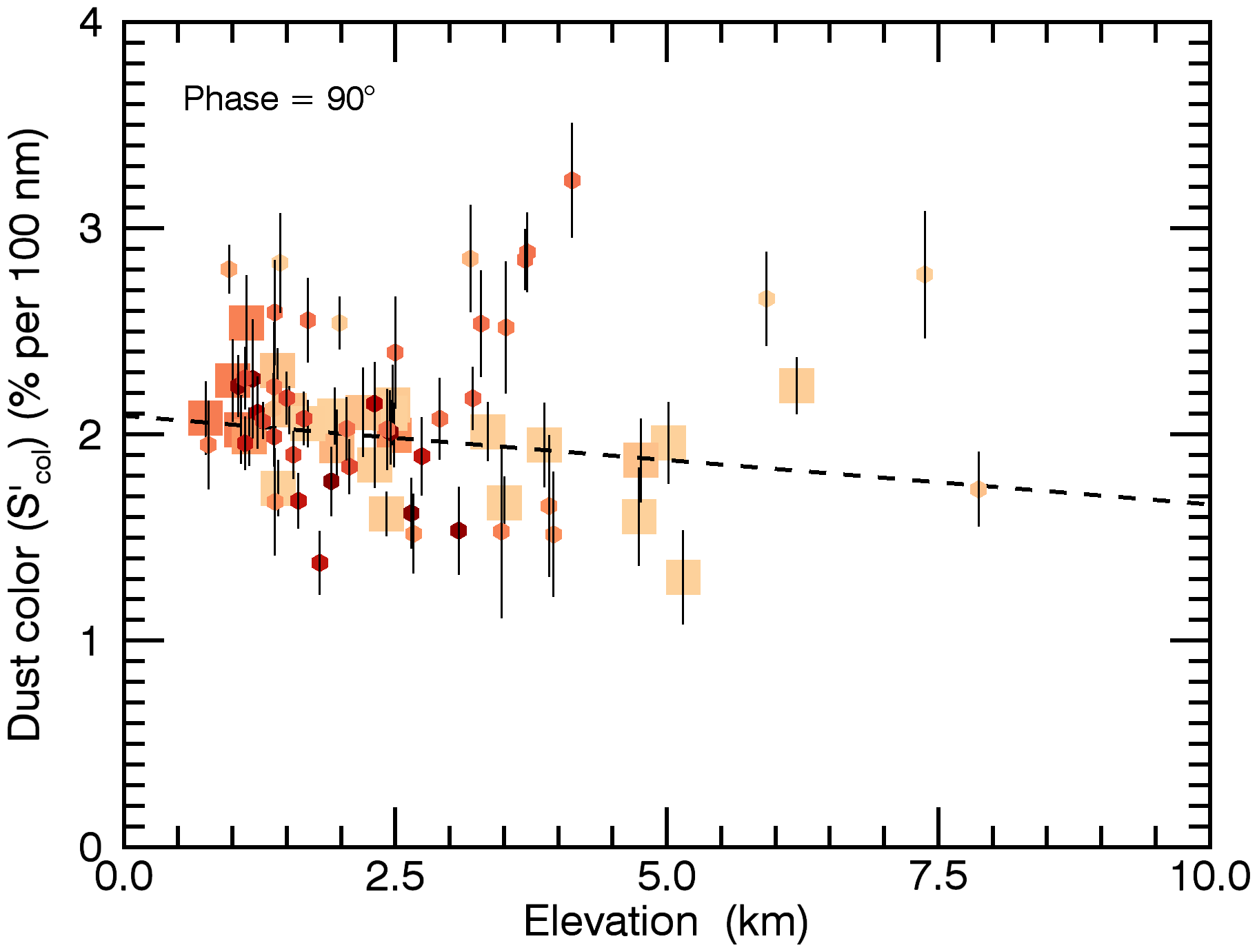}
\end{minipage}    
       \caption{Variation of bolometric albedo, superheating factor and color with elevation $H$. Data obtained with phase angle $\alpha$ = 83--90$^{\circ}$ are considered. The color is a function of the heliocentric distance, as given by the colorbar. Only data with {\it TEST}$_{3.0}$ $<$ 1.35 are plotted. Those with $TEST_{3.0}$ $<$ 1.1 are shown with large squares. The dashed lines correspond to a power law (for albedo ($\propto$ $H^{-0.39\pm0.01}$)) or a linear fit (for superheating factor and color) to the data points with $TEST_{3.0}$ $<$ 1.1.}
     \label{fig:fit-PH90}
\end{figure}

In this Section, we only consider measurements obtained at phase angles between  83$^{\circ}$ and 90 $^{\circ}$ (mean value of 89$^{\circ}$). These data were acquired mainly before perihelion. The color temperature follows $T_{\rm col}$ = (338 $\pm$ 1)$r_{\rm h}^{-0.60 \pm 0.01}$ K in the heliocentric range $r_{\rm h}$ 1.24--1.5 AU. Considering only the best quality data (covering 1.24--1.34 AU), one finds $T_{\rm col}$ = (333 $\pm$ 3)$r_{\rm h}^{-0.51 \pm 0.03}$ K.

\begin{figure}
\includegraphics[width=\columnwidth]{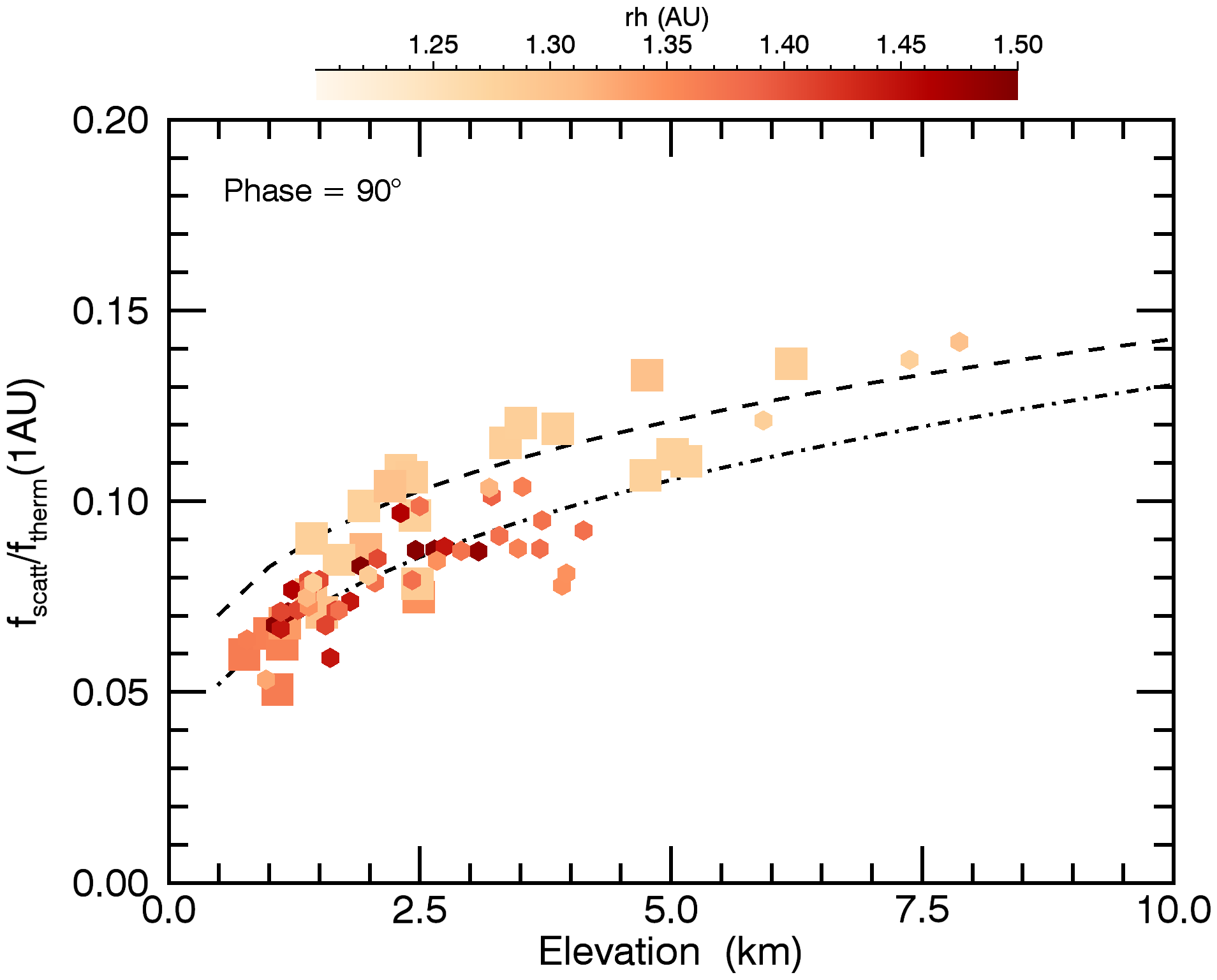}    
 \caption{Variation of $f_{\rm scatt}/f_{\rm therm}$(1AU) (deduced from the ratio of the radiances at 2.44 $\mu$m  and 4.6 $\mu$m, see text) with elevation $H$ for $\alpha$ = 83--90$^{\circ}$. The color coding and symbols are as for Fig.~\ref{fig:fit-PH90}.  Data with {\it TEST}$_{3.0}$ $<$ 1.35 are considered.
The dashed-dotted and dashed lines correspond to a power law fit for data obtained  
between --71 d and  --10 d wrt perihelion ($\propto$ $H^{+0.31}$) and  between --2 d to 21 d ($\propto$ $H^{+0.22}$), respectively.  }
     \label{fig:fluxratio-PH90}
\end{figure}

Figure~\ref{fig:fit-PH90} shows the bolometric albedo, color, and superheating factor 
as a function of elevation $H$ (and $r_{\rm h}$ using a color gradient for the symbols).  The results from the highest quality spectra ($TEST_{3.0}$ $<$ 1.1) are shown with squares, whereas the other data (1.1 $<$ $TEST_{3.0}$ $<$ 1.35) are shown with dots. $S_{\rm heat}$ and $S'_{\rm col}$ have mean values of 1.19$\pm$0.01 and 2.0$\pm$0.2 
\% per 100 nm, respectively. Lower quality spectra show lower $S_{\rm heat}$ and higher color $S'_{\rm col}$ and albedo values that may be inaccurate (see Sect.~\ref{sec:modelfitting}). To test this hypothesis, we performed spectral fitting,  fixing the color temperature. We found that an underevaluation of $S_{\rm heat}$ by 4\% ($S_{\rm heat}$=1.15 instead of 1.2) would decrease the derived albedo by $\sim$60\%. Effectively the albedo derived for the low quality spectra giving $S_{\rm heat}$=1.15 is lower by this order of magnitude (upper panel of Fig.~\ref{fig:fit-PH90}). So, results from these spectra, especially those for which the derived color $S'_{\rm col}$ is well above the mean value, are a priori doubtful. On the other hand, the intensity ratio 
$f_{\rm scatt}/f_{\rm therm}$(1AU), which is proportional to the bolometric albedo, presents a similar behavior with elevation and heliocentric distance though discrepancies between high quality and low quality spectra are somewhat smaller  (Fig.~\ref{fig:fluxratio-PH90}).
In conclusion, the trend for an enhanced albedo at low heliocentric distance seen in Fig.~\ref{fig:fit-PH90} is likely real, as well as the trend for increased superheating with decreasing $r_{\rm h}$.

A marginal decrease of $S_{\rm heat}$ and $S'_{\rm col}$ with increasing elevation $H$ is suggested (best data), with a Pearson correlation coefficient $R$ of --0.34 and --0.40, respectively (Fig.~\ref{fig:fit-PH90}). We performed a multi regression analysis for studying variations with both $r_{\rm h}$ and altitude. A weak $r_{\rm h}$ variation in $r_{\rm h}^{-0.15 \pm 0.05}$ is suggested for $S_{\rm heat}$, which improves the correlation coefficient with altitude to $R$ = --0.55, with $S_{\rm heat}$ $\propto$ $H^{-0.009\pm0.003}$.
 Multi regression analysis did not provide convincing results for $S'_{\rm col}$:  no reliable variation of the color with  $r_{\rm h}$ could be identified in this data set. Altogether, however, variations of $S_{\rm heat}$ and $S'_{\rm col}$ with $H$ and $r_{\rm h}$ (1.24 to 1.35 AU) are small.
 
\begin{figure}
    \includegraphics[width=\columnwidth]{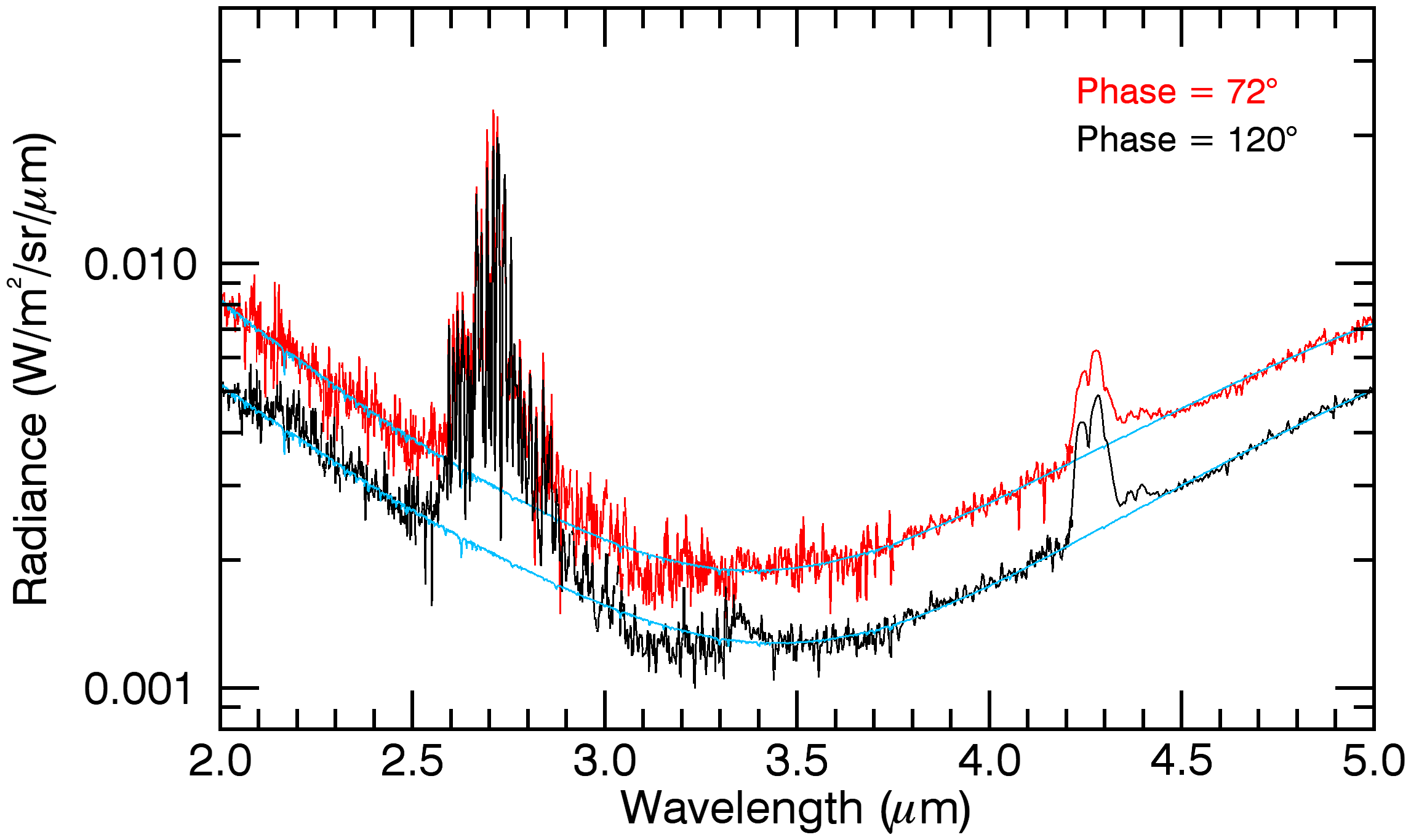}
 \caption{VIRTIS-H spectra of comet 67P obtained at different phase angles (shown in logarithmic scale). Top spectrum (red): cube T1$\_$00399392471 acquired on 28 Aug. 2015 ($r_{\rm h}$ = 1.26 AU, $\alpha$= $\sim$72$^\circ$) with a mean elevation of 4.1 km. Bottom spectrum (black): cube T1$\_$00400433767 acquired on 7 Sept. 2015 ($r_{\rm h}$ = 1.29 AU, $\alpha$= $\sim$120$^\circ$) with a mean elevation of 2.2 km. The model fits to the continuum are shown in cyan, with derived parameters ($T_{\rm col}$(K), $A$, $S'_{\rm col}$(\% per 100 nm), $S_{\rm heat}$) = (304, 0.14, 1.3, 1.23) and (283, 0.09, 2.8, 1.15) for 28 Aug. and 7 Sept., respectively. The spectra fulfill the quality criterion $TEST_{3.0}$ $<$ 1.1 and do not present significant stray light.
 }
     \label{fig:sp-low-high-phase}
\end{figure}  
 
 There is evidence for a significant increase of the bolometric albedo with $H$ (Fig.~\ref{fig:fit-PH90}). This is illustrated in Fig.~\ref{fig:splowhighalbedo}, which displays two spectra obtained at $H$= 1.4 and 6.2 km, the former showing a lower flux ratio $f_{\rm scatt}/f_{\rm therm}$. Since $S_{\rm heat}$ (or $T_{\rm col}$) shows weak variation with $H$, the increase of $A$ with $H$ reflects the increase of $f_{\rm scatt}/f_{\rm therm}$(1AU)  with $H$, shown in Fig.~\ref{fig:fluxratio-PH90}. 
We looked for possible variations of $A$ with $r_{\rm h}$ or seasonal changes, performing a multi regression analysis to $f_{\rm scatt}/f_{\rm therm}$(1AU). Comparing data acquired between --2 d to 21 d wrt perihelion to those acquired before (up to end July 1015), an average increase of 20\% of $f_{\rm scatt}/f_{\rm therm}$(1AU) (and hence of the albedo) is suggested (Fig.~\ref{fig:fluxratio-PH90}). The variation with elevation follows 
$f_{\rm scatt}/f_{\rm therm}$(1AU) $\propto$ $H^{+0.27\pm0.05}$, where the power law index is the average of the indexes obtained for the two time periods (Fig.~\ref{fig:fluxratio-PH90}). The bolometric albedo measured on the high quality spectra follows the same variation.

\subsection{Phase variations} 

The dust color and color temperature exhibit a strong correlation with phase angle. The dust color is larger at large phase angles   (Figs~\ref{fig:res_allphase}C). On the other hand, the reverse is observed for the color temperature, as best seen when looking to the trend followed by the superheating factor (Fig.~\ref{fig:res_allphase}D). Figure~\ref{fig:sp-low-high-phase} compares two spectra acquired with a one week interval at $\alpha$= 72 and 120$^\circ$. The ratios of the thermal emissions in orders 1 (3.7--4.2 $\mu$m) and 0 (4.5--5 $\mu$m) present subtle differences (by up to 9\%) explained by a color temperature higher by 20 K at low phase. The fitting algorithm retrieves also a bluer color at low phase to match the 3.0--3.5 $\mu$m radiances.

\begin{figure}[H]
\centering
\begin{minipage}{8cm}
    \includegraphics[width=\columnwidth]{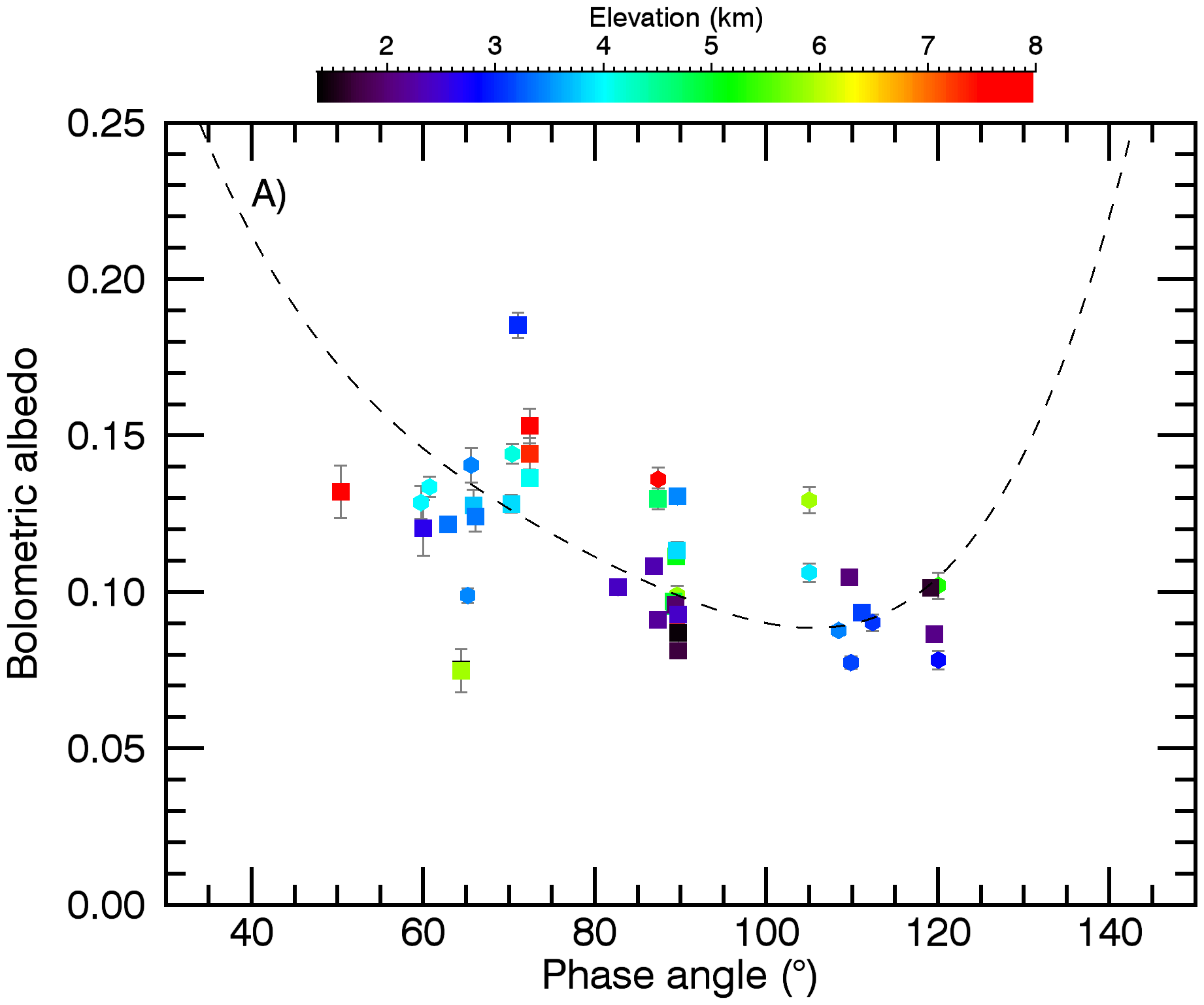}
\end{minipage}
\begin{minipage}{8cm}
       \includegraphics[width=\columnwidth]{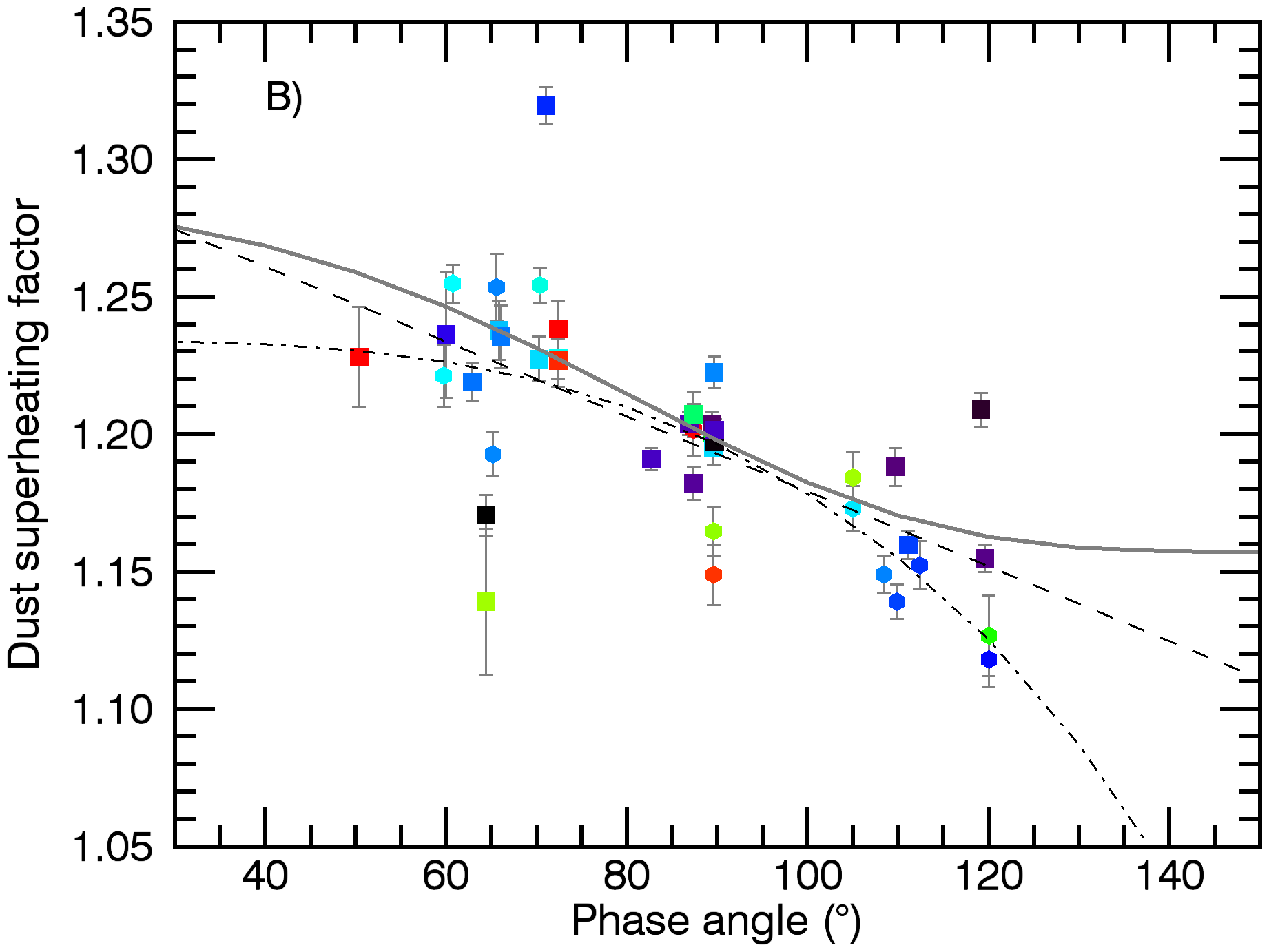}       
\end{minipage}
\begin{minipage}{7.2cm}    
    \includegraphics[width=7.6cm]{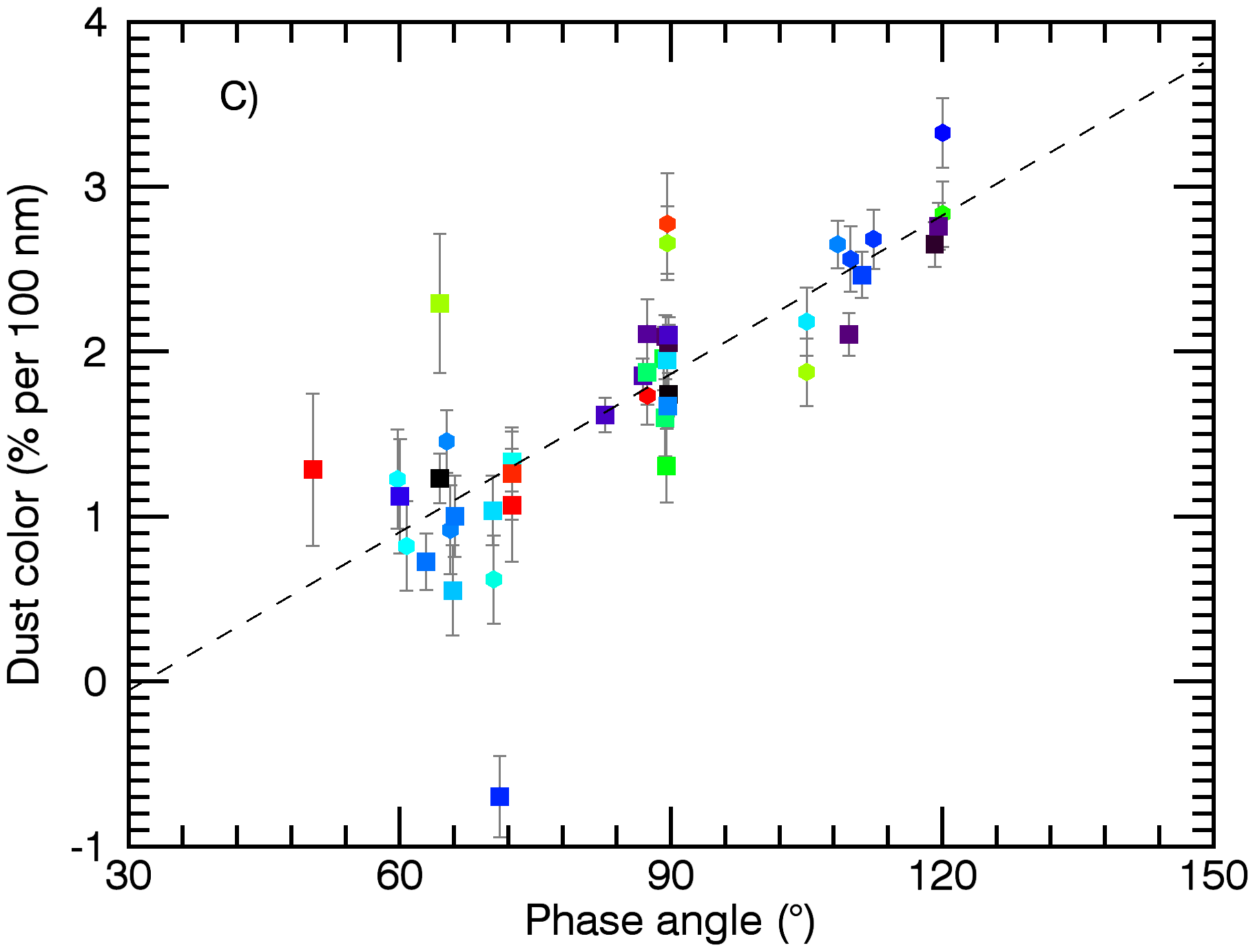}
\end{minipage}    
       \caption{Variation of bolometric albedo $A$ (top, A), superheating factor $S_{\rm heat}$ (middle, B) and color $S'_{\rm col}$ (bottom, C) with phase angle using data taken after 11 August 2015. Data are from spectra with {\it TEST}$_{3.0}$ $<$ 1.35 with the best data ($TEST_{3.0}$ $<$ 1.1) shown with squares. A) The dashed line is the phase function measured during MTP020 by \citet{Bertini2017} (OSIRIS Green filter -- 537 nm). B) The dashed line is a linear fit to $S_{\rm heat}$ data with a slope of --0.0014/$^\circ$ (--0.33 K/$^\circ$ at $r_{\rm h}$ = 1.35 AU). The dashed-dotted line is from the Near Earth Asteroid Thermal Model   (NEATM) with beaming parameter $\eta$ = 1.6, emissivity $\epsilon$ = 0.9 and $r_{\rm h}$ = 1.35 AU. The solid line is from our dust thermal model (Sec.~\ref{sec:discu-temp}) with thermal parameter $\Theta$ = 0.1, optical depth fraction  of isothermal particles $f_{\rm iso}$ = 0.8, $f_{\rm heat}$ = 1.05, and $\epsilon$ = 0.9.  C) The dashed line is a linear fit to the data giving a phase reddening of 0.032 \%/100 nm/$^\circ$.  }
     \label{fig:fit-allphase}
\end{figure}

We present in Figs.~\ref{fig:fit-allphase}B and C the variations of the color and superheating factor with phase angle. To avoid clutter at $\alpha$ = 90$^\circ$, only dates after --2 d wrt perihelion are plotted. The phase dependences found using the best quality data are $\sim$ 0.3 K/$^\circ$ for $T_{\rm col}$, and 0.031 \%/100 nm/$^\circ$ for the dust color. Significant variations with elevation are not seen.

The bolometric albedo (measured at 2 $\mu$m) follows a phase variation which matches the phase function measured at 537 nm by \citet{Bertini2017} during  MTP020/STP071 (end August 2015)  (Fig.~\ref{fig:fit-allphase}A). The VIRTIS data present a large scatter, which prevents further comparison. Note that the dust phase function is expected to be wavelength-dependent.  The variation of bolometric albedo with elevation at low phases ($\alpha$ < 80$^\circ$) follow a $H^{0.25}$ law for the best data,  similar to the one  measured at $\alpha$ = 90$^\circ$, but the data show significant scatter with respect to this variation.   

Phase variations of color and color temperature of cometary dust have never been reported in the literature. From detailed analysis and multiple checks, we can rule out biases related to the fitting algorithm and data quality. Since retrieved parameters are somewhat correlated (Sect.~\ref{sec:modelfitting}), another test was to fix the color temperature according to Eq.~\ref{eq:1}, with $S_{\rm heat}$ fixed to the $\alpha$ = 90$^{\circ}$ value of 1.19 (Sect.~ \ref{sec:terminator}). Despite the increase of the $\chi^2$ values, the color trend with phase remains. However, the bolometric albedo shows a monotonic slight decrease with decreasing phase angle (i.e., no backscattering enhancement), which is unexpected from scattering models,   
thus reassuring us that the observed phase variations are real.

\section{Discussion}
\label{sec:discussion}

In summary, analysis of the dust 2--5 $\mu$m continuum radiation from 67P's coma shows: 
i) a mean dust color of 2\%/100 nm and superheating factor of 1.19 at 90$^\circ$
phase angle, consistent with previous VIRTIS-H measurements \citep{dbm2017}; ii) a factor 2.5 increase of the bolometric albedo with increasing elevation from $H$ = 0.5 to 8 km; iii) an increase of dust color temperature with decreasing phase angle at the rate of $\sim$ 0.3 K/$^\circ$ ; iv) spectral phase reddening at a rate of 0.032 \%/100 nm/$^\circ$. More marginally, decreasing color temperature and color with increasing $H$ are possibly observed, as well as 20\% higher albedo values after perihelion.  

\subsection{Phase reddening}
\label{sec:discu-color}

The photometric properties of the dust coma present similarities with the nucleus surface. 67P's nucleus shows a phase reddening which has been observed both in the optical (VIS) (0.5--0.8 $\mu$m) and in the near-IR (1--2 $\mu$m) ranges \citep{Ciarniello2015,Longobardo2017,Feller2016,Fornasier2016}. In the near IR, 67P's nucleus color is 3.9 \%/100 nm at $\alpha$ = 90$^\circ$, with a phase reddening  between 0.013--0.018 \%/100 nm/$^\circ$  \citep{Ciarniello2015,Longobardo2017}. Phase reddening is higher in the VIS (0.04 to 0.1 \%/100 nm/$^\circ$), with lower values near perihelion associated to a bluing of the surface \citep{Fornasier2016}.  For the dust coma, the weighted mean of the VIS values measured by \citet{Bertini2017} using OSIRIS data (excluding spurious MTP026 results) yields  0.025\%/100 nm/$^\circ$. This is close to the values that we are measuring in the near-IR.  However, it should be kept in mind that the VIS values are from data with LOS perpendicular to the nucleus--S/C vector \citep{Bertini2017}, so they pertain to the dust coma in the near-spacecraft environment, whereas the near-IR values characterise the near-nucleus coma. There are several lines of evidence that the dust properties vary with elevation, as discussed later on.

Phase reddening is observed for many Solar System bodies, including zodiacal light \citep{Leinert1981}.
For planetary surfaces, phase reddening can be interpreted as an effect of multiple scattering. For dark and porous bodies as 67P, multiple scattering is relevant despite the low albedo thanks to the increase of scattering surfaces caused by the roughness of the particles present on the nucleus surface \citep{Schroder2014}. Laboratory experiments combined with numerical simulations  have indeed highlighted the role of microscopic roughness in producing such a spectral effect \citep{Beck2012,Schroder2014}. Particle irregularities at a spatial scale less than the wavelength are also invoked to explain the phase reddening seen in the visual for interplanetary dust (10--100 $\mu$m sized)  \citep{Schiffer1985}. Then, the phase reddening observed in the 67P coma could be related to the porous structure of the particles, providing those contributing to scattered light are sufficiently large. The relative similarity in the phase curves of the dust coma and surface (especially the backscattering enhancement) is consistent with the predominance of large and fluffy dust particles in the coma, as, e.g., discussed by \citet{Moreno2018}, \citet{Bertini2019} and \citet{Markkanen2018}. Other evidence for relatively large ($\geq$ 10  $\mu$m) scatterers in the coma of 67P include dust tail modeling \citep{Moreno2017} and the unexpected low amount of submicron and micron-sized particles collected by the Rosetta's MIDAS experiment \citep{Mannel2017}. 

\subsection{Phase variation of the color temperature}
\label{sec:discu-temp}

The color temperature excess with respect to the equilibrium temperature expected for isothermal grains is a common property of cometary atmospheres. The superheating factor measured for 67P of $\sim$ 1.2 is in the mean of values observed in other comets \citep{dbm2017}.  This temperature excess is usually attributed to the presence of submicrometric grains composed of absorbing material \citep{Hanner2003,Kolokolova2004}. \citet{dbm2017} showed that this temperature excess could result from the contribution of hot fractal-like aggregates to near-IR thermal emission, these particles having in turn little input to scattered light. In this case, based on Mie modeling, the minimum size of the more numerous and more compact particles would be $\geq$ 20 $\mu$m \citep{dbm2017}. The observed decrease of the color temperature with increasing phase angle can not be explained by variations of the dust size distribution with solar azimuth angle \citep{Shou2017}, which would induce a phase curve symmetric with respect to $\alpha$ = 90$^{\circ}$. On the other hand, this trend can be caused by non-isothermal grains showing day-to-night thermal contrast. This explanation holds for Saturn's C-ring whose thermal emission shows variations with solar phase angle \citep{Altobelli2008,Leyrat2008}.   

To test this hypothesis, in a first approach we used the Near Earth Asteroids Thermal Model (NEATM) \citep{Harris1998} for describing the variation of the temperature over the surface of comet dust particles. NEATM assumes an idealized non-rotating spherical object with a temperature decreasing from a maximum at the subsolar point to zero at the terminator (there is no night-side emission). For low albedo bodies, the surface temperature at latitude $\theta'$--$\pi$/2 and longitude $\phi'$ (subsolar point at $\theta'$ = 90$^{\circ}$ and $\phi'$ = 0$^{\circ}$) follows:

\begin{equation}
T_{\rm NEATM} = T_{\rm NEATM}^{\rm SS} (\sin(\theta') \cos(\phi'))^{0.25},
\label{eq:eq2}
\end{equation}

\noindent
with

\begin{equation}
T_{\rm NEATM}^{\rm SS} = \frac{394}{r_{\rm h}^{0.5} (\eta\epsilon)^{0.25}} [K],
\label{eq:eq2a}
\end{equation}

\noindent 
where $\epsilon$ is the emissivity (taken equal to 0.9), and $\eta$ is the so-called beaming parameter, which is used in asteroid studies as a calibration coefficient to account for the effects of thermal inertia, rotation and surface roughness. $T_{\rm NEATM}^{\rm SS}$ is the temperature at the subsolar point. Thermal emission is calculated considering the surface elements facing the observer and, therefore, depends on the phase angle \citep{Harris1998}.  We computed NEATM 3--5 $\mu$m spectra for a range of phase angles and $\eta$ values. By fitting a blackbody to these spectra, we derived color temperatures and, using Eq.~\ref{eq:1}, the corresponding superheating factors. The phase variation of these computed superheating factors (dashed-dotted curve in Fig.~\ref{fig:fit-allphase}B) matches the variation measured for 67P dust, and the mean observed value $S_{\rm heat}$ = 1.19 at $\alpha$ $\sim$ 90$^{\circ}$ (Sect.~\ref{sec:terminator}) is obtained for $\eta$ = 1.58. We determined $\eta$ for each of the data points shown in Fig.~\ref{fig:fit-allphase}B. Inferred $\eta$ values do not show significant phase dependence and average out at 1.59 $\pm$ 0.17. 
This value is intermediate between the limiting cases $\eta$ = 1 (high day-to-night contrast due to low 
thermal inertia, slowly spinning particles, or spin axis along Sun direction) and $\eta$ = 4 (isothermal particles). This suggests that both isothermal and non-isothermal grains are contributing to 67P dust thermal emission in the 3--5 $\mu$m wavelength range.

\begin{figure}
\includegraphics[width=\columnwidth]{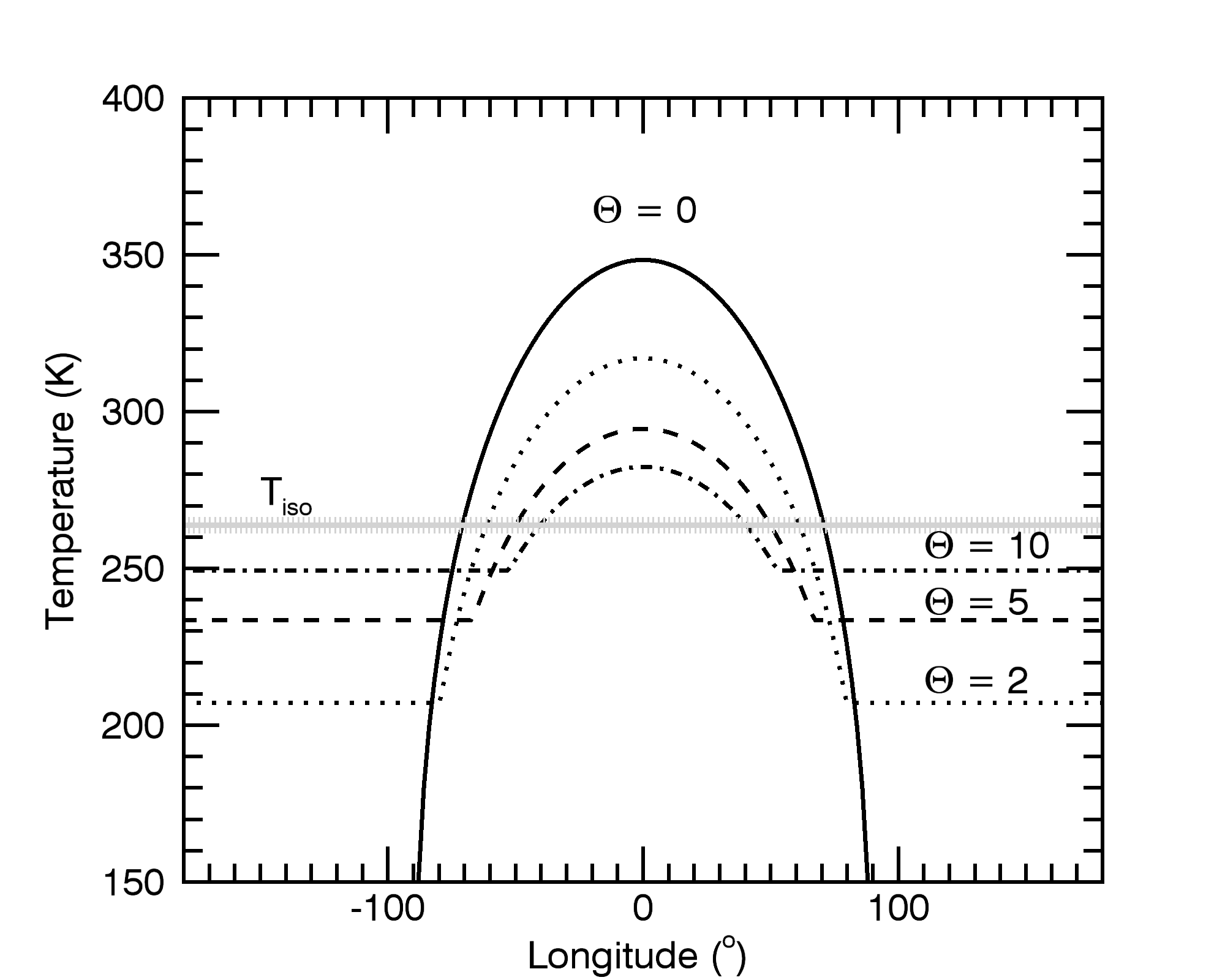}
\caption{Modelled particles diurnal temperature profiles at equator for different values of the thermal parameter $\Theta$. The subsolar point is at latitude zero and at longitude zero. For large $\Theta$, the temperature reaches the constant isothermal value $T_{\rm iso}$. Calculations are for $r_{\rm h}$ = 1.35 AU and $\epsilon$=0.9. }
\label{fig:temperature_profile}

\end{figure}

\begin{figure}

\centering
\begin{minipage}{9cm}
       \includegraphics[width=\columnwidth]{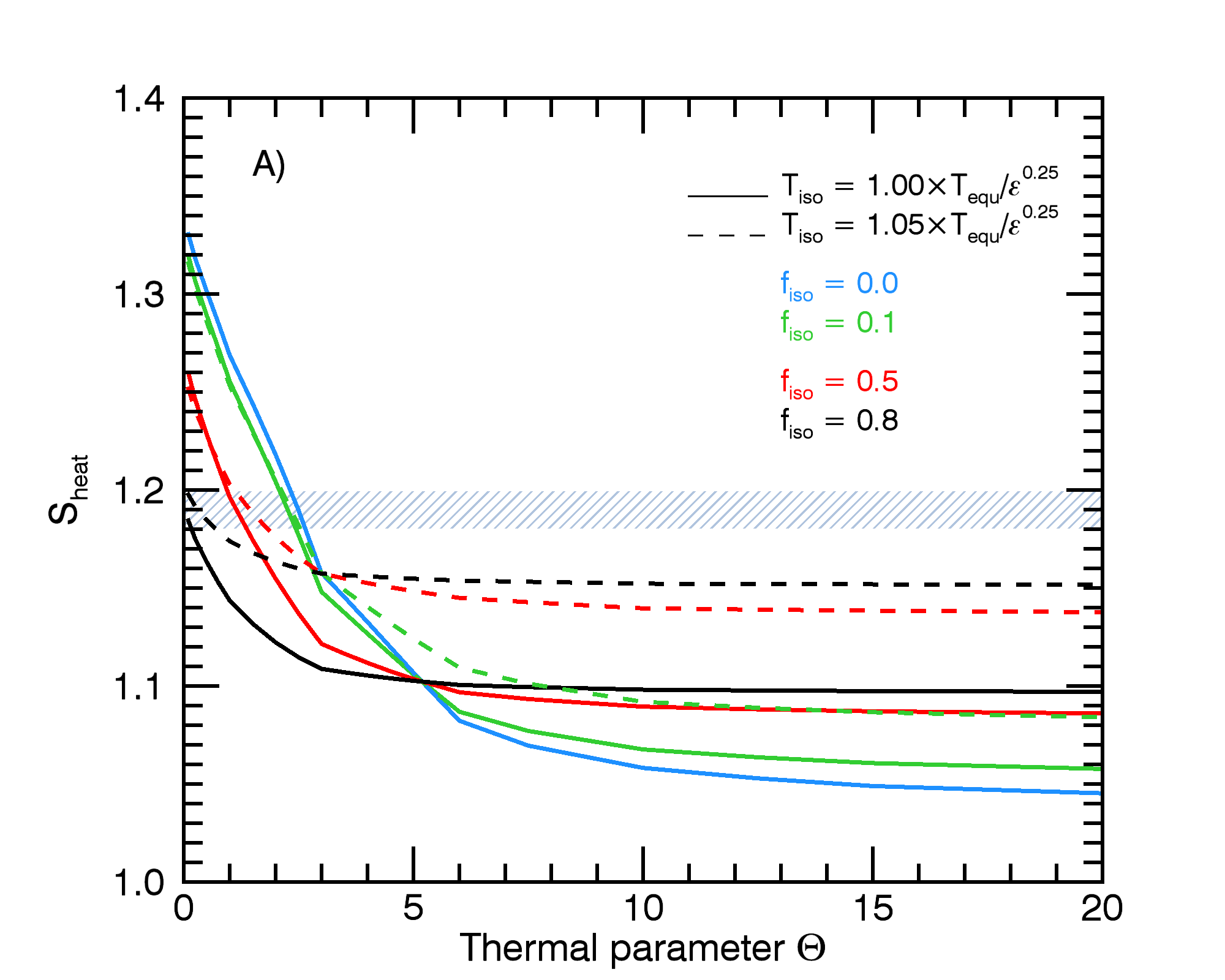}  
\end{minipage} 
\begin{minipage}{9cm}  \vspace{-0.1cm}        
       \includegraphics[width=\columnwidth]{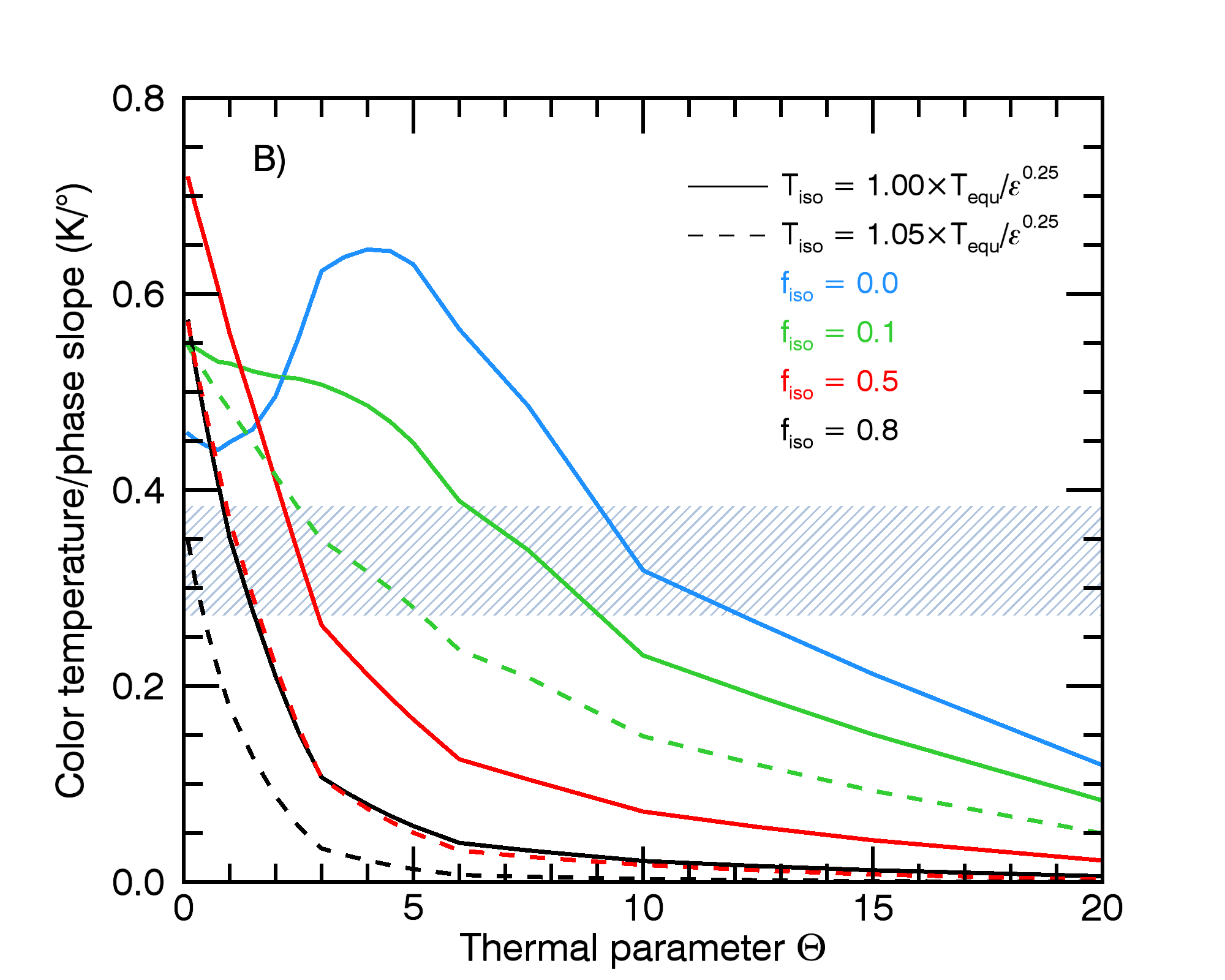}      
\end{minipage}
\caption{Superheating factor $S_{\rm heat}$ (A) and phase linear dependence of the color temperature (B) from the thermal model parameterized by the thermal parameter $\Theta$, the optical depth contribution of isothermal particles $f_{\rm iso}$ and their
physical temperature $T_{\rm iso}$. The shaded areas correspond to the measurements in the 67P coma.} 
\label{fig:thermal-result}
\end{figure}

To go further into the interpretation of the data, we developed a simple model (Appendix~\ref{sec:appendixB}), considering a bimodal distribution of dust particles consisting of a mixture of isothermal particles and particles presenting day-to-night temperature contrast. The diurnal temperature profile of non-isothermal particles is described by the {\it thermal parameter} $\Theta$ introduced by \citet{Spencer1989}, which depends on their thermal properties (which is a function of porosity) and spinning rate. Figure~\ref{fig:temperature_profile} shows examples of diurnal temperature profiles, which computations are described in Appendix~\ref{sec:appendixB}. Expected $\Theta$ values for 67P dust particles are also given in Appendix~\ref{sec:appendixB} (Fig.~\ref{fig:thermal-prop}). The relative contribution of the isothermal particles to the total optical depth is parameterized by the quantity $f_{\rm iso}$ (in the range 0--1), and their physical temperature $T_{\rm iso}$ is in excess with respect to the equilibrium temperature by a factor $f_{\rm heat}/\epsilon^{0.25}$.  Figure~\ref{fig:thermal-result} shows the superheating factor and the slope of the phase variation of the color temperature as a function of $\Theta$ for different values of $f_{\rm iso}$, considering values of $f_{\rm heat}$ of 1.0 and 1.05. The non-monotonic behavior of the phase dependence for low $\Theta$ and $f_{\rm iso}$ values is because the 3--5 $\mu$m wavelength range is more sensitive to high temperatures (e.g., for high day-to-night temperature contrast, only the warm surface areas contribute to the brightness). Best match to the measurements is obtained for $\Theta$ $\leq$ 2, corresponding to a significant day-to-night temperature contrast ($>$ 1.5). Such low values of $\Theta$ imply slowly spinning particles with high porosity, low thermal inertia or non-spinning particles (Fig.~\ref{fig:thermal-prop}). The relative contribution of isothermal particles is not heavily constrained. For $\Theta$ = 2.0, one finds solutions with $f_{\rm iso}$ = 0.2--0.4. On the other hand, for $\Theta$ = 0.1, a good fit to the data is obtained for  $f_{\rm iso}$ = 0.8 (see Fig.~\ref{fig:fit-allphase}). We note that a good fit to the color temperature and its phase variation is obtained providing the physical temperature of the isothermal grains is in excess by 8\% with respect to the expected equilibrium temperature (i.e., $f_{\rm heat}$ = 1.05, considering an assumed emissivity of 0.9). Therefore, the presence of non-isothermal grains with day-side surface temperature well above the equilibrium temperature, cannot alone explain the superheating factor observed for cometary dust. As already discussed, a possible explanation is a significant contribution of submicron-sized absorbing grains \citep{Hanner2003} or, alternatively, of highly porous fractal-like aggregates with sub-micron size monomers, as these particles can be warmer than more compact particles  \citep{dbm2017}.

A realistic size distribution of the dust particles is obviously not bimodal. It is interesting to estimate the critical radius below which the particles are isothermal, and to compare it to estimated diurnal skin depths (Appendix~\ref{sec:appendixB}, Fig.~\ref{fig:thermal-prop}). Assuming a power law for the size distribution (d$N$ $\propto$ a$^{-\beta}$d$a$, where $a$ is the particle radius), this critical radius $a_{\rm crit}$ depends on the size index $\beta$ and minimum and maximum sizes of the particles,  $a_{\rm min}$ and $a_{\rm max}$, and can be computed using the inferred relative contribution to the total optical depth of the two populations of particles ($f_{\rm iso}$ and 1--$f_{\rm iso}$) \citep[see equations in][]{Leyrat2008}. $a_{\rm crit}$ increases with increasing $a_{\rm min}$, $a_{\rm max}$ and  $f_{\rm iso}$, and with decreasing $\beta$. Let us consider size ranges $a_{\rm min}$ = 1--20 $\mu$m and $a_{\rm max}$ = 1--10 cm, consistent with constraints obtained for 67P dust \citep{dbm2017,Mannel2017,Ott2017,Schloerb2017,Moreno2018,Markkanen2018}. 
For $f_{\rm iso}$ = 0.8 (solution obtained for $\Theta$ = 0.1) and  $\beta$ = 2.5 (respectively $\beta$ = 3.0), $a_{\rm crit}$ is in the range 0.6 to 6 cm (respectively 0.15--1.7 cm). These values of $a_{\rm crit}$ are on the order of or larger than the estimated diurnal skin depths of $\sim$ 0.3 cm for slowly spinning  and high porosity particles with $\Theta$ = 0.1 (Appendix~\ref{sec:appendixB}, Fig.~\ref{fig:thermal-prop}).
For $f_{\rm iso}$ = 0.3 (solution obtained for $\Theta$ = 2.0), $a_{\rm crit}$ is in the range 0.09 to 0.9 cm (respectively 0.0015--0.02 cm) for $\beta$ = 2.5 (respectively $\beta$ = 3.0). For particles with porosity of 0.5--0.9 and spinning rates consistent with $\Theta$ = 2.0, we expect diurnal skin depths from 0.01 to 0.3 cm. Altogether, except for size distributions with the opacity dominated by small particles (those with $\beta$ = 3 and $a_{\rm min}$ $<$ 10 $\mu$m,  or $\beta >$ $3$), we infer that the 
critical particle size separating isothermal and non-isothermal particles is on the order or larger than the diurnal skin depth. This is a satisfactory result since we expect particles with size smaller than the  diurnal skin depth to be isothermal due to internal heat transfer.

\subsection{Radial variation of the bolometric albedo}

Measured bolometric albedos for 67P quiescent dust coma range between 0.05 and 0.15 at 90$^{\circ}$ phase angle (Fig.~\ref{fig:fit-PH90}) and encompasse values measured in other comets, as previously discussed by \citet{dbm2017}. These values correspond to a low geometric albedo, and is consistent with dust particles made of dark material \citep[][and references therein]{Kolokolova2004,dbm2017}.  
The VIRTIS-H observations suggest an increase of the dust bolometric albedo with increasing radial distance (Sect.~\ref{sec:results}). Albedo maps obtained for comets 1P/Halley and 21P/Giacobini-Zinner by combining visible light and thermal infrared images show a similar trend:  
the albedos increase radially from the nucleus, except along the tail, where the albedos are smaller \citep{Telesco1986,Hammel1987}. 
Variations in albedo may result from different composition, particle size, shape and structure. For example, large fluffy grains may have reduced albedos because they induce multiple scattering events that allow more light to be absorbed. For this reason, the lower albedos near the nucleus and in the tail of comets 1P and 21P have been interpreted as due to the presence of large, fluffy grains escaping the nucleus with low velocities  and confined in the orbital planes of the comets \citep{Telesco1986,Hammel1987}. We may thus invoke an enhanced proportion of chunks in the inner coma of 67P, in line with
the conclusion obtained by \citet{Bertini2019} from the variation of the backscattering enhancement with nucleocentric distance. During the perihelion period, comet 67P underwent numerous outbursts \citep{Vincent2016}, which likely populated the inner coma with large, slowly-moving dust particles, as observed for comet 17P/Holmes after its massive 2007 outburst \citep{Reach2010,Boissier2012}. In addition, evidence for particles falling back to the nucleus are several \citep{Keller2017}. Models of the density distribution for a coma dominated by gravitationally bound particles on ballistic trajectories predict an excess of particles in the inner coma with respect to the density expected for free-radial outflow \citep{Chamberlain1987,Gerig2018}. There are some hints of such a deviation from free-radial outflow in OSIRIS optical images \citep{Gerig2018}, which would be amplified if the observed trend for a smaller albedo at smaller cometocentric distances is considered. Deviations are also conspicuous for the dust thermal radiation measured in the microwave, which samples essentially large particles, and shows a steep decrease of the column density at impact parameters below 10 km \citep{Schloerb2017}.   

Whereas we argued previously for radial variations of the optical properties of the individual grains, changes in the particle size distribution may also affect the bolometric albedo of the coma. Anomalously high bolometric albedos were measured in the very active comet C/1995 O1 (Hale-Bopp), and during strong jet activity of 1P/Halley, which were found to be correlated with a high silicate 10$\mu$m-band contrast, and a high superheating factor $S_{\rm heat}$, suggesting that the presence of a large amount of small particles was responsible for these high albedos \citep[][]{Tokunaga1986,Mason2001,Hanner2003}. Similarly, the rapidly moving 67P outburst ejecta displayed high $A$ and $S_{\rm heat}$, together with blue colors, characteristics of small particles \citep{dbm2017}. Mie calculations for a porous mixture of olivine and amorphous carbon at 90$^{\circ}$ phase angle predict an increase of $A$ from a value of 0.05, when only particle sizes $>$ 1 $\mu$m are considered, to values up to 0.20 when submicron particles are present. However, the increase of $A$ is expected to be correlated with an increase of the superheating factor. This trend between $A$ and $S_{\rm heat}$ with increasing elevation is not observed (Sect.~\ref{sec:results}, Fig.~\ref{fig:fit-PH90}). Therefore, dust fragmentation is likely not responsible for the increase of $A$ with elevation. 

Changes in the albedo may also be related to a change in the particle composition. Particles made of less absorbing material are expected to be brighter, cooler and bluer. This trend is observed with increasing elevation, which would then imply that evaporation of some dark material took place in the inner coma. Evidence for the degradation of grains in the coma of 67P are however still very sparse \citep[e.g., hydrogen halides and glycine are released from dust,][]{Dekeyser2017,Altwegg2016}. Incidentally, we note that, in presence of rapidly subliming (i.e., small and dirty) ice grains, the trend would have been opposite. 

The VIRTIS-H observations suggest an increase by $\sim$ 20\% of the bolometric albedo in the -2 d to 21 d wrt perihelion period, when the comet was the most active, possibly associated with an increase of $S_{\rm heat}$. This trend would be in line with an increased number of small particles at perihelion time, or alternatively, with enhanced degradation of dark material. The 67P nucleus surface showed a global enhancement of water ice content near perihelion  \citep{Ciarniello2016,Fornasier2016}. The observed $A$ increase would be in line with an expected increased amount of icy grains in the inner coma of 67P. On the other hand, this does not explain the trend observed for $S_{\rm heat}$.

\section{Summary and conclusion}

Spectra of the dust 2--4.5 $\mu$m continuum radiation have been acquired with the VIRTIS-H experiment aboard the Rosetta mission to comet 67P. Through spectral fitting, we measured the dust color temperature, bolometric albedo and 2--2.5 $\mu$m color. From the analysis of data acquired from 3 June to 29 Oct. 2015  ($r_{\rm h}$ = 1.24--1.55 AU) at line-of-sight tangent altitudes between 0.5 and 10 km, the following results were obtained:

\begin{itemize}

\item At phase angles $\sim$90$^{\circ}$, the color temperature varied from 260 K to 320 K, and followed a $r_{\rm h}^{-0.6}$ law, close to the $r_{\rm h}^{-0.5}$ variation expected from the balance between absorbed solar radiation and radiated thermal energy. A 20\% increase of the bolometric albedo is observed near perihelion.

\item A mean dust color of 2\%/100 nm and superheating factor of 1.19 are measured at 90$^\circ$ phase angle, consistent with previous VIRTIS-H measurements \citep{dbm2017}.

\item A decrease of the color temperature with increasing phase angle is observed, at a rate of $\sim$ 0.3 K/$^\circ$. It can be explained by the presence of large porous particles, with low thermal inertia, presenting a significant day-to-night temperature contrast.
 
\item A large spectral phase reddening is measured. The value (0.032 \%/100 nm/$^\circ$) is higher then values measured for the nucleus in the near IR \citep[0.013--0.018 \%/100 nm/$^\circ$,][]{Ciarniello2015,Longobardo2017}. This phase reddening can be related to the roughness of the dust particles. 

\item The bolometric albedo was found to increase from 0.05 to 0.14 (i.e., by a factor 2.5) with increasing tangent altitude (so-called elevation in the paper) from 0.5 to 8 km. A decrease of the color temperature and color with increasing altitude is marginally observed. Possible explanations include the presence in the inner coma of dark particles on ballistic trajectories, and changes in particle composition.  

\item Evidence for grain fragmentation, or disappearance of icy grains, are not seen.

\end{itemize}  

In future papers, we seek exploring the infrared continuum images obtained with VIRTIS-H to obtain further constraints on the dust coma of comet 67P. 


\section*{Acknowledgements}
The authors would like to thank the following institutions and
agencies,  which supported this work: Italian Space Agency (ASI -
Italy), Centre National d'Etudes Spatiales (CNES -- France),
Deutsches Zentrum f\"{u}r Luft- und Raumfahrt (DLR -- Germany),
National Aeronautic and Space Administration (NASA -- USA). VIRTIS
was built by a consortium from Italy, France and Germany, under
the scientific responsibility of the Istituto di Astrofisica e
Planetologia Spaziali of INAF, Rome (IT), which lead also the
scientific operations.  The VIRTIS instrument development for ESA
has been funded and managed by ASI, with contributions from
Observatoire de Meudon financed by CNES and from DLR. The
instrument industrial prime contractor was former Officine
Galileo, now Leonardo company in Campi Bisenzio,
Florence, IT. The authors wish to thank the Rosetta Science Ground
Segment and the Rosetta Mission Operations Centre for their
fantastic support throughout the early phases of the mission. The
VIRTIS calibrated data shall be available through the ESA's
Planetary Science Archive (PSA) Web site. With fond memories of
Angioletta Coradini, conceiver of the VIRTIS instrument, our
leader and friend. D.B.M. thanks E. Lellouch for enlightening discussions.

\begin{appendix}

\section{}
\label{sec:appendixA}

\begin{table}[h]
    \caption{Log of VIRTIS-H observations and retrieved dust spectral and temperature properties.
    (\textit{This Table is not available in the present version.)}}
    \label{tab:logtable}
\end{table}

\section{Thermal model for dust particles}

\label{sec:appendixB}
A simple approach, inherited from asteroid studies, is used to model the particle surface temperature as a function of local time. \citet{Spencer1989} showed that, for smooth objects with the Sun in the equatorial plane, the diurnal temperature profile can be parameterized by a quantity $\Theta$ called {\it thermal parameter}. $\Theta$ is
function of the thermal inertia $\Gamma$, angular rotation rate $\omega$, subsolar equilibrium temperature $T_{\rm SS}$, emissivity $\epsilon$, and Stefan-Boltzmann constant $\sigma$:

\begin{equation}
\Theta = \frac{\Gamma \sqrt{\omega}}{\epsilon \sigma T_{\rm SS}^3}, 
\end{equation}

\noindent
with

\begin{equation}
\Gamma = \sqrt{K C_p^*},
\end{equation}

\noindent
where $K$ is the thermal conductivity and  $C_p^*$ is the volumetric heat capacity.

The subsolar equilibrium temperature for dark and smooth objects is given by:

\begin{equation}
T_{\rm SS} = \frac{394}{r_{\rm h}^{0.5} (\epsilon)^{0.25}} [K].
\label{eq:TSS}
\end{equation}

Objects with higher $\Theta$ will have lower day-to-night temperature contrast, lower subsolar temperature and maximum temperature in the afternoon.  Objects with low thermal inertia and low rotation rate have low $\Theta$, so high temperature contrasts with maximum temperature near midday.  

The diurnal skin depth is given by \citep{Legall2014}: 
 
\begin{equation}
\delta_{\rm therm} = \frac{\Gamma}{C_p^*} \sqrt{\frac{2}{\omega}}.
\end{equation}

\noindent
 The surface becomes isothermal with time of the day and depth for $\Theta$ $>$ 50 \citep{Spencer1989}. However, diurnal temperature profiles parameterized by $\Theta$ \citep{Spencer1989} are invalid for objects with sizes in the range or lower than the diurnal skin depth $\delta_{\rm therm}$. Due to 
solar heat penetration, they should be isothermal. We computed $\Theta$ and  
$\delta_{\rm therm}$ for a range of parameters appropriate to 67P porous dust aggregates (Table~\ref{tab:param}). The thermal conductivity $K_{\rm mat}$, specific heat capacity $C_p$, and volumetric density $\rho_{\rm mat}$ of the material composing the monomers (i.e., subunits) correspond to values measured for carbonaceous meteorites. The adopted $\rho_{\rm mat}$ value is also consistent with estimations of the compacted bulk density of 67P dust particles \citep{Fulle2017}. The volumetric heat capacity and thermal conductivity of the aggregates depend on their porosity $\phi$. We have:  
 
\begin{equation}
C_p^* = C_p (1-\phi) \rho_{\rm mat}.
\end{equation}

For the thermal conductivity, we used the empirical formula derived by \citet{Arakawa2017} for highly porous dust aggregates, giving:   

\begin{equation}
K = c_{\rm k} K_{\rm mat} (1-\phi)^2.
\end{equation}

The coefficient of proportionality $c_{\rm k}$ is given in \citet{Arakawa2017}.  Comparing with laboratory measurements, \citet{Arakawa2017} found that this porosity dependence can be used for $\phi$ $> 0.5$. The thermal conductivity of aggregates derived from this empirical formula is $K$ = 0.00714 to 0.000285 W m$^{-1}$ K$^{-1}$  for $\phi$ = 0.5 to 0.9.

Rotation frequencies ($\omega$/2$\pi$)  of 67P grains have been estimated by dust dynamical simulations \citep{Fulle2015,Ivanovski2017a,Ivanovski2017b}, and also measured using the light curves of individual dust particles \citep{Fulle2015}. Computed values for $>$ 100 $\mu$m size particles range from 0.03--4 Hz depending of grain characteristics (shape, size and density), gas flow environment and initial conditions \citep{Fulle2015,Ivanovski2017a}. Measurements in the low activity phase of the comet show that the most probable rotation frequency of the grains is below 0.15 Hz \citep{Fulle2015}. The simulations show that particles experience oscillations before acquiring a full rotation at a few kilometers from the surface, with the rotation axis perpendicular to the flow direction. Modeling the surface temperature of oscillating grains is beyond the scope of this paper. 

\begin{figure}

\centering
\begin{minipage}{9cm}
    \includegraphics[width=\columnwidth]{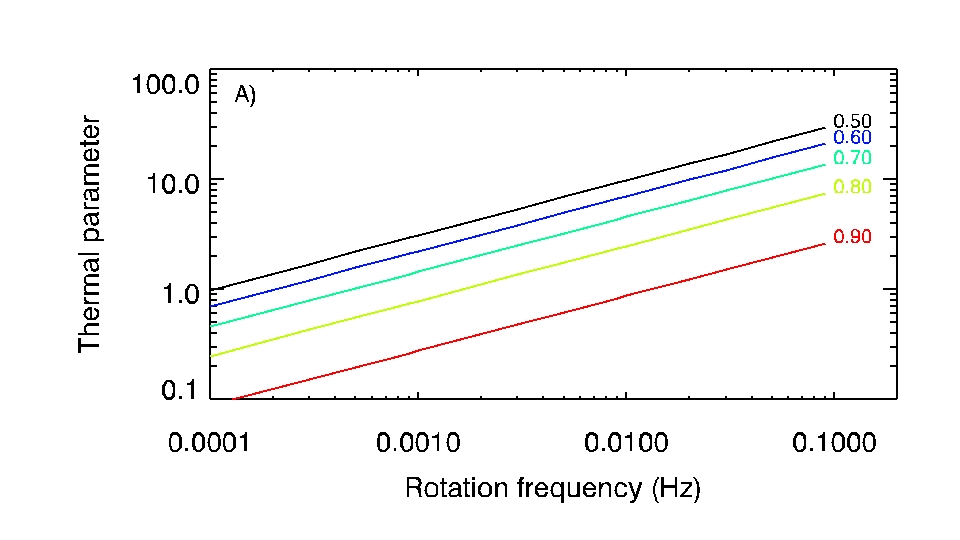}
 \end{minipage} 
\begin{minipage}{9cm}  \vspace{-0.3cm}   
    \includegraphics[width=\columnwidth]{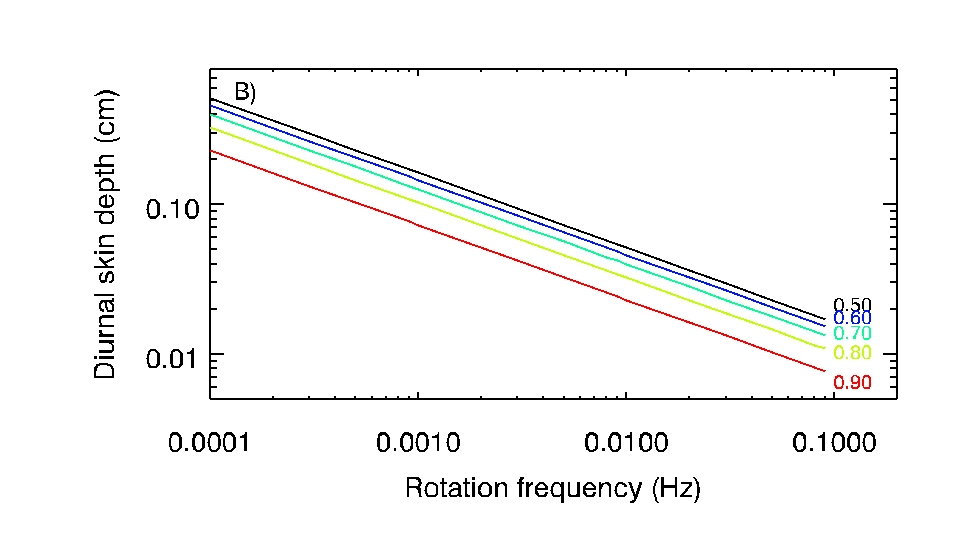}       
\end{minipage}
\begin{minipage}{9cm}  \vspace{-0.3cm}   
    \includegraphics[width=\columnwidth]{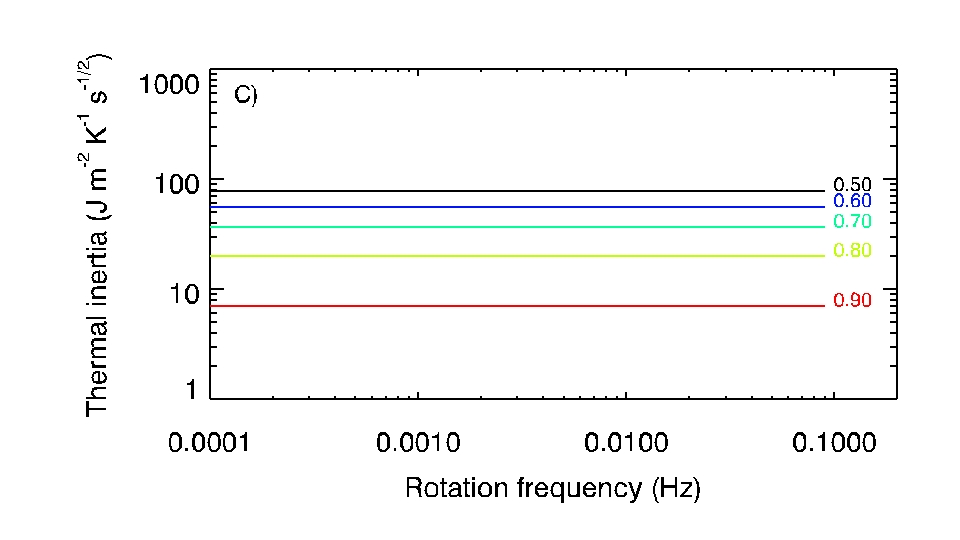}       
\end{minipage}
\caption{Thermal parameter $\Theta$, diurnal skin depth $\delta_{\rm therm}$ and thermal inertia $\Gamma$ as a function of particle rotation frequency. The different curves correspond to different porosities, with values indicated on the plots. }\label{fig:thermal-prop}
\end{figure}

The thermal parameter, diurnal skin depth and thermal inertia for porosities of 0.5--0.9 and rotation frequencies of 0.0001--0.1 Hz are shown in Fig.~\ref{fig:thermal-prop}. Calculations were performed with the parameters given in Table~\ref{tab:param}. Thermal inertias are low (8--80 J m$^{-2}$ K$^{-1}$ s$^{-1/2}$). The diurnal skin depth ranges from $\sim$ 0.01 to 0.5 cm, and the thermal parameter varies from less than 1 (strong day-to-night contrast) for slowly rotating grains to values $\sim$ 30 for relatively compact aggregates ($\phi$ = 0.5) with rotation frequency of 0.1 Hz.

\citet{Spencer1989} provide diurnal temperature profiles on the equator as a funtion of $\Theta$.
In their Fig. 2, they show the variation of $T_{\rm MAX}$/$T_{\rm SS}$, $T_{\rm MIN}$/$T_{\rm SS}$ with $\Theta$, where $T_{\rm MAX}$ and $T_{\rm MIN}$ are respectively the maximum and minimum surface temperatures, and $T_{\rm SS}$ the subsolar temperature for a non-rotating body (Eq.~\ref{eq:TSS}). We described the surface temperature at latitude $\theta'$--$\pi$/2 and longitude $\phi'$ (subsolar point at equator, i.e.,  $\theta'$ = 90$^{\circ}$ and $\phi'$ = 0$^{\circ}$) by:


\begin{align}
\label{eq:A1}
&T_{day}(\theta',\phi',\Theta) = \nonumber\\
&\max\Big(T_{\rm MAX}(\Theta) (\sin(\theta') \cos(\phi'))^{0.25},T_{night}(\theta',\phi',\Theta)\Big)
\end{align}

\begin{equation}
\label{eq:A2}
T_{night}(\theta',\phi',\Theta) = T_{\rm MIN}(\Theta)\sin(\theta')^{0.25},
\end{equation}

\noindent
where $\theta'$ is in the range (0, $\pi$), and $\phi'$ is in the range (--$\pi$,  $\pi$). This ($\theta'$, $\phi'$) dependence corresponds to NEATM for $\Theta$ = 0, and should approximate well the temperature distribution when $\Theta$ $>$ 0, except that the shift of the maximum temperature in the afternoon for nonzero $\Theta$ is not represented. $T_{\rm MAX}(\Theta)$ and $T_{\rm MIN}(\Theta)$ were taken from  \citet{Spencer1989}. Figure~\ref{fig:temperature_profile} shows diurnal temperature curves for different $\Theta$.

We then computed the 3--5 $\mu$m thermal emission as a function of phase angle, considering the surface elements facing the observer (cf. NEATM model from \citet{Harris1998}). We used a bimodal distribution for the grains, consisting of  isothermal grains at $T$=$T_{\rm iso}$ (e.g., rapidly spinning dust particles, or grains with size smaller than $\delta_{\rm therm}$), and slowly spinning/low thermal inertia particles, with a temperature profile described by $\Theta$ (Eq~\ref{eq:A1}--\ref{eq:A2}). This approach follows the one adopted by \citet{Leyrat2008} to explain the phase variation of the color temperature of Saturn's C ring. The relative contribution of isothermal particles is given by $f_{\rm iso}$, which determines the optical depth contribution of isothermal particles ($f_{\rm iso}$ = $\tau_{\rm iso}$/($\tau_{\rm iso}+\tau_{\rm non-iso}$)). The optical depth is proportional to the integral over size range of the size distribution times particle cross-section \citep[e.g.][]{Leyrat2008}. The temperature of the isothermal particles is expressed as:

\begin{equation}
T_{\rm iso} = \frac{f_{\rm heat} T_{\rm equ}}{\epsilon^{\rm 0.25}},
\end{equation}

\noindent
where $T_{\rm equ}$, which corresponds to the equilibrium temperature for an emissivity equal to 1, is given in Eq.~ \ref{eq:1}. The parameter $f_{heat}$ allows us to investigate dust particles heated above the equilibrium temperature. This temperature excess is expected for small particles made of absorbing material \citep[][and references therein]{Kolokolova2004}. 

Table~\ref{tab:paramfree} summarizes the free parameters of the model for computing synthesized spectra of dust thermal emission. By fitting a blackbody function to these spectra, the superheating factor as a function of phase angle can be derived and compared to VIRTIS-H measurements.

\begin{table}
\caption{Model fixed parameters.\label{tab:param}}
\begin{tabular}{l l l l}        
\hline\hline \noalign{\smallskip}
Parameter & & Value & Note \\
\hline \noalign{\smallskip}
{\it Aggregate subunits:} & & & \\
Thermal conductivity & $K_{\rm mat}$ & 0.7$^a$ &  CM chondrites \\
(W/m/K) & & \\
Heat capacity & $C_{\rm p}$ & 700$^b$ & 300 K -- Chondrites\\
(J/kg/K) & & \\
Density  & $\rho_{\rm mat}$ & 2420$^c$ & Orgueil\\
(kg/m$^3$) & & \\
{\it Aggregate} & & \\
Emissivity & $\epsilon$ &0.9 & \\
\hline \noalign{\smallskip}
\end{tabular}

{\footnotesize
$^a$ \citet{Opeil2010}
$^b$ \citet{Consolmagno2013}
$^c$ \citet{Macke2011}
}
\end{table}

\begin{table}
\caption{Model free parameters.\label{tab:paramfree}}
\begin{tabular}{l l  }        
\hline\hline \noalign{\smallskip}
Parameter & Description \\
\hline \noalign{\smallskip}
$\alpha$ & Phase angle \\
$\Theta$ & Thermal parameter \\
$f_{\rm iso}$ & Optical depth contribution \\
& of isothermal particles \\
$f_{\rm heat}$ & Factor to equilibrium temperature \\
& of isothermal particles \\
\hline \noalign{\smallskip}
\end{tabular}
\end{table}

\end{appendix}

\end{document}